\documentclass[a4paper,11pt]{article}
\pdfoutput=1
\usepackage{jcappub_nt}
\usepackage[utf8]{inputenc}

\usepackage{bm}

\title{Primordial gravitational waves and perturbations during an inhomogeneous inflation}

\author{Xian Gao}

\author{and Chao Kang}

\affiliation{School of Physics and Astronomy,\\ 
	Sun Yat-sen University, Guangzhou 510275, China}

\emailAdd{gaoxian@mail.sysu.edu.cn}

\emailAdd{kangch@mail2.sysu.edu.cn} 

\abstract{
	We investigate the inhomogeneous inflation, in which the space exponentially expands with inhomogeneities, and its cosmological perturbations.
	The inhomogeneous inflation is realized by introducing scalar fields with spacelike gradients that break the spatial symmetry.
	We find that the space can expand uniformly in different direction with the same rate.
	By using the perturbative method, we calculate the corrections to the power spectra of gravitational waves and curvature perturbation up to the linear order in the background inhomogeneities.
	Since the background is inhomogeneous, perturbations modes with different wave numbers get correlated.
	We show that generally the power spectra of perturbations depend on the ratio and the angle of wave numbers of the two correlated modes.
	In particular, the two circular polarization modes of the gravitational waves gain different powers when the background inhomogeneity is of vector or tensor type.
	}


\arxivnumber{1804.03652}

\begin{document}

\maketitle


\section{Introduction}

The accelerating expansion of our universe in its primordial epoch is one of the mysteries in modern cosmology \cite{Sato:2015dga}.
There are two questions which are related to each other.
One is the question of initial conditions for inflation.
Precisely, if inflation --- although itself will rapidly inflate away any classical inhomogeneities, would occur with spatially inhomogeneous initial conditions.
Although such problems have not yet been completely settled down, there have been some researches recently which shown that inflation is rather robust to the inhomogeneous initial conditions \cite{Goldwirth:1990pm,Berezhiani:2015ola,Easther:2014zga,East:2015ggf,Clough:2016ymm} (see \cite{Brandenberger:2016uzh} for a short review of initial conditions for inflation).

We would not address ourselves to the question of robustness of inflation to inhomogeneous initial conditions.
Instead, this work is devoted to answering the other question: whether the universe would have inflated with an inhomogeneous spatial sector. 
The aim of this paper is thus two-fold.
On the one hand, we have to examine whether the exponential expansion of the  spatial sector with an inhomogeneous geometry is indeed one of the solutions to the equations of motion. 
One may view the standard inflation scenario with homogeneous spatial sector as the homogeneous inflation, and thus dub such an exponential expansion of the inhomogeneous spatial sector  as the \emph{inhomogeneous inflation}.
On the other hand, we would like to investigate whether such an inhomogeneous inflation has impacts on the observables, i.e., the power spectra of the primordial perturbations.

Comparing with the standard homogeneous inflation, such an inhomogeneous inflation scenario receives much less attention.
First it is difficult to realize such an inhomogeneous inflationary background with a single matter content, e.g., in the single scalar-field inflation models. 
Moreover, one may expect naively that any matter content
that could cause the spatial inhomogeneities would also affect
the expansion rates in different spatial directions, giving rise to the increase of spatial anisotropy. 
If this is the case, such inhomogeneous inflation scenarios are definitely ruled out by the current astrophysical observations.
Nevertheless, a solution with uniform expansion of the inhomogeneous spatial sector was found in \cite{Geng:2014vza}, in which the spatial isotropy --- in the sense that  different spatial directions expand with the same rate, is preserved.
Other than the usual inflaton field with spatially homogeneous background value, the key ingredients in \cite{Geng:2014vza} are scalar-fields with spatially inhomogeneous but time-independent background values, which are responsible to the inhomogeneous but uniformly expanding space geometry.

The idea of having matter fields with time-independent but space-dependent background configurations was firstly introduced in the ``elastic inflation'' \cite{Gruzinov:2004ty} and further systematically developed in ``solid inflation'' \cite{Endlich:2012pz}, where the scalar fields are Goldstone bosons associated with breaking of spatial diffeomorphism.
Similar to the usual effective field theory of inflation \cite{Dubovsky:2005xd,Creminelli:2006xe,Cheung:2007st}, where the adiabatic mode of matter perturbations is associated with the breaking of time diffeomorphism, the effective field theory of spacetime symmetry breaking was also developed \cite{Leutwyler:1996er,Son:2005ak,Nicolis:2013lma,Hidaka:2014fra,Nicolis:2015sra,Bartolo:2015qvr,Lin:2015cqa}.

Perturbations respect the symmetry of the background they live.
As a result, one may expect that  power spectra of perturbations on the spatially inhomogeneous background would no longer take the standard form, i.e., $\sim \delta^3(\bm{k}+\bm{k}') P(k)$.
This is also the case of perturbations in the anisotropic inflation models (see e.g. \cite{Soda:2012zm} for a review), where correlation
functions of perturbations show statistically anisotropic features \cite{Ackerman:2007nb,Gao:2009vi,Watanabe:2010fh,Barnaby:2012tk,Bartolo:2012sd,Shiraishi:2013vja,Shiraishi:2013oqa,Emami:2013bk,Abolhasani:2013zya,Emami:2014tpa,Ashoorioon:2015pia,Ashoorioon:2016lrg}.
Cosmological perturbations in solid inflation and its variations have also been widely studied \cite{Bartolo:2013msa,Bartolo:2014xfa,Sitwell:2013kza,Ricciardone:2016lym,Celoria:2017bbh,Endlich:2013jia,Cannone:2014uqa,Cannone:2015rra,Koh:2013msa,Kouwn:2014aia,Namjoo:2014nra,Jazayeri:2014nya,Akhshik:2014gja,Firouzjahi:2016fxf}.
In our model, the number of spacelike scalar fields is not fixed to three, which is required in the solid inflation.
In particular, our model generalizes the solid inflation in the sense that we do not require a homogeneous spatial geometry, and thus will have much more fruitful observational features.

This paper is organized as following.
In Sec.\ref{sec:formalism} we set up our general formalism, and make the ansatz for the inhomogeneous background configurations. 
Then we develop linear perturbation theory around this inhomogeneous inflation background.
In Sec.\ref{sec:gw} we calculate the corrections to the power spectra of gravitational waves, by assuming the background inhomogeneity is small.
In Sec.\ref{sec:sp} we calculate the corrections to the power spectrum of scalar (curvature) perturbation.
Sec.\ref{sec:con} concludes.

We set $M_{\mathrm{pl}} \equiv (8\pi G_{\mathrm{N}})^{-1} =1$.

\section{The formalism} \label{sec:formalism}

Let us consider the following action
\begin{equation}
S=\int\mathrm{d}^{4}x\sqrt{-g}\left[\frac{1}{2}R-\frac{1}{2}\left(\partial\Phi\right)^{2}-V\left(\Phi\right)-\frac{1}{2}g^{\mu\nu}f_{IJ}\partial_{\mu}\phi^{I}\partial_{\nu}\phi^{J}\right], \label{model}
\end{equation}
where $f_{IJ}$ is the metric in field space of $\phi^I$ with $I,J=1,\cdots,\mathcal{N}$.
In terms of ADM variables, the physical metric is written as
\begin{equation}
\mathrm{d}s^{2}=-N^{2}\mathrm{d}\tau^{2}+g_{ij}\left(\mathrm{d}x^{i}+N^{i}\mathrm{d}\tau\right)\left(\mathrm{d}x^{j}+N^{j}\mathrm{d}\tau\right),\label{metric_ADM}
\end{equation}
where $N^{i}=g^{ij}N_{j}$ and $\tau$ is the comoving time.
We define the background to be 
\begin{equation}
\mathrm{d}s^{2}=a^{2}\left(-\mathrm{d}\tau^{2}+\bar{h}_{ij}\mathrm{d}x^{i}\mathrm{d}x^{j}\right), 
\label{bg_h}
\end{equation}
where $a=a(\tau)$ is the usual scale factor, $\bar{h}_{ij}=\bar{h}_{ij}\left(\vec{x}\right)$ is the ``background'' spatial metric which we assume to time-independent.
The metric (\ref{bg_h}) describes an expanding universe with inhomogeneous spatial sector.

The ADM variables $\left\{ N,N_{i},g_{ij}\right\} $ are parametrized by
\begin{eqnarray}
N & = & ae^{A},\label{N}\\
N_{i} & = & a^{2}B_{i},\label{Ni}\\
g_{ij} & = & a^{2}h_{ij}=a^{2}\left(\bar{h}_{ij}+H_{ij}\right),\label{gij}
\end{eqnarray}
where $A$, $B_i$ and $H_{ij}$ denote the deviation from the background metric (\ref{bg_h}).
Similarly, the scalar fields are also splitted into the background parts and perturbations as
\begin{eqnarray}
\Phi\left(\tau,\vec{x}\right) & = & \bar{\Phi}\left(\tau\right)+\delta\Phi\left(\tau,\vec{x}\right), \label{bg_Phi}\\
\phi^{I}\left(\tau,\vec{x}\right) & = & \bar{\phi}^{I}\left(\vec{x}\right)+\delta\phi^{I}\left(\tau,\vec{x}\right). \label{bg_phiI}
\end{eqnarray}
Similar to the usual inflaton field, the background value of $\Phi$ is time-dependent and spatially homogeneous.
On the other hand, the background values of $\phi^I$ are time-independent and spatially inhomogeneous, which is the key ansatz in this work. 
For later convenience, we may further split $B_i$ and $H_{ij}$ into irreducible parts
\begin{eqnarray}
B_{i} & = & \partial_{i}B+\tilde{B}_{i},\label{Bi_dec}\\
H_{ij} & = & 2\zeta\,\bar{h}_{ij}+2\left(\bar{\mathrm{D}}_{i}\bar{\mathrm{D}}_{j}-\frac{1}{3}\bar{h}_{ij}\bar{\mathrm{D}}^{2}\right)E+2\bar{\mathrm{D}}_{(i}F_{j)}+\gamma_{ij},\label{Hij_dec}
\end{eqnarray}
where $\bar{\mathrm{D}}_{i}$ is the covariant derivative adapted to $\bar{h}_{ij}$ and $\bar{\mathrm{D}}_{i}\tilde{B}^{i}=\bar{\mathrm{D}}_{i}F^{i}=\bar{\mathrm{D}}_{i}\gamma^{ij}=\gamma_{\phantom{i}i}^{i}=0$.
Here and in what follows, spatial indices are raised and lowered by $\bar{h}_{ij}$ and its inverse $\bar{h}^{ij}$.

\subsection{The inhomogeneous background} \label{sec:bg}

Let us verify that the above background configurations (\ref{bg_h}), (\ref{bg_Phi}) and (\ref{bg_phiI}) are indeed solutions of our model. The background equations of motion are determined by requiring the vanishing of the first order Lagrangian for perturbations, which reads
\begin{equation}
S_{1} = \int\!\mathrm{d}\tau\mathrm{d}^{3}x\sqrt{\bar{h}}\,a^{2}\left(\mathcal{E}^{(A)}A+\frac{1}{2}\mathcal{E}_{ij}^{(H)}H^{ij}-\mathcal{E}^{(\Phi)}\delta\Phi+\bar{f}_{IJ}\mathcal{E}^{(\phi)I}\delta\phi^{J}\right),
\end{equation}
with $\bar{h} \equiv \det \bar{h}_{ij}$ and
\begin{eqnarray}
\mathcal{E}^{(A)} & \equiv & 3\mathcal{H}^{2}-\frac{1}{2}\bar{\Phi}'^{2}-a^{2}\bar{V}+\frac{1}{2}\bar{\mathcal{R}}-\frac{1}{2}\bar{h}^{ij}\bar{f}_{ij}=0,\label{bgeom_A}\\
\mathcal{E}_{ij}^{(H)} & \equiv & \left(2\frac{a''}{a}-\mathcal{H}^{2}+\frac{1}{2}\bar{\Phi}'^{2}-a^{2}\bar{V}-\frac{1}{2}\bar{h}^{kl}\bar{f}_{kl}\right)\bar{h}_{ij}-\bar{\mathcal{G}}_{ij}+\bar{f}_{ij}=0,\label{bgeom_H}\\
\mathcal{E}^{(\Phi)} & \equiv & \frac{1}{a^{2}}\left(a^{2}\bar{\Phi}'\right)'+a^{2}\bar{V}_{,\Phi}=0,\label{bgeom_Phi}\\
\mathcal{E}^{(\phi)I} & \equiv & \bar{\nabla}^{2}\bar{\phi}^{I}+\partial_{i}\bar{\phi}^{K}\partial^{i}\bar{\phi}^{L}\bar{\Gamma}_{KL}^{I}=0,\label{bgeom_phi}
\end{eqnarray}
where $\mathcal{H}\equiv a'/a$ is the comoving Hubble parameter and we defined
\begin{equation}
\bar{f}_{ij}\equiv\bar{f}_{IJ}\partial_{i}\bar{\phi}^{I}\partial_{j}\bar{\phi}^{J},
\end{equation}
for short, a ``$'$'' denotes derivative with respect to $\tau$, $\bar{\mathcal{R}}$ and $\bar{\mathcal{G}}_{ij}$ are Ricci scalar and Einstein tensor of the background spatial metric $\bar{h}_{ij}$, $\bar{\Gamma}_{KL}^{I}$ is the Christoffel symbol of $\bar{f}_{IJ}$.
It is interesting that all the equations above are naturally separately into the temporal parts and spatial parts thanks to the fact that $\bar{h}_{ij}$ is time-independent.
For example, $\mathcal{E}^{(A)}=0$ implies the following two independent equations
\begin{equation}
3\mathcal{H}^{2}-\frac{1}{2}\bar{\Phi}'^{2}-a^{2}\bar{V}=-\frac{1}{2}\left(\bar{\mathcal{R}}-\bar{h}^{ij}\bar{f}_{ij}\right)\equiv-\frac{3}{2}\mathcal{K}, \label{bgeom_A_sp}
\end{equation}
where $\mathcal{K}$ must be some numerical constant (of dimension $[M]^2$).
We are free to choose $\mathcal{K}=0,\pm 1$ due to the scaling symmetry.
For the uniform expansion of the scale factor $a(\tau)$, a non-vanishing $\mathcal{K}$ plays the role of a cosmological constant.
Similarly, the trace part of $\mathcal{E}^{H}_{ij}=0$ can be written as
\begin{equation}
2\frac{a''}{a}-\mathcal{H}^{2}+\frac{1}{2}\bar{\Phi}'^{2}-a^{2}\bar{V}=-\frac{1}{6}\left(\bar{\mathcal{R}}-\bar{h}^{ij}\bar{f}_{ij}\right)\equiv-\frac{1}{2}\mathcal{K}. \label{bgeom_Htr_sp}
\end{equation}
On the other hand, the traceless part of  $\mathcal{E}^{H}_{ij}=0$ is given by
	\begin{equation}
	\frac{1}{3}\bar{h}_{ij}\left(\bar{\mathcal{R}}-\bar{h}^{kl}\bar{f}_{kl}\right)-\bar{\mathcal{R}}_{ij}+\bar{f}_{ij}=0,
	\end{equation}
together with $\bar{\mathcal{R}}-\bar{h}^{ij}\bar{f}_{ij}=3\mathcal{K}$ it implies
\begin{equation}
\bar{\mathcal{R}}_{ij}=\bar{f}_{ij}+\mathcal{K}\bar{h}_{ij}. \label{Rij_fin}
\end{equation}
As usual, combining the temporal parts of (\ref{bgeom_A_sp}) and (\ref{bgeom_Htr_sp}) yields
\begin{equation}
\mathcal{H}'-\mathcal{H}^{2}+\frac{1}{2}\bar{\Phi}'^{2}=\frac{1}{2}\mathcal{K}. \label{bgeom_tc}
\end{equation}
For later convenience, we define
\begin{equation}
\epsilon\equiv1-\frac{\mathcal{H}'}{\mathcal{H}^{2}},\qquad\tilde{\epsilon}\equiv\frac{\bar{\Phi}'^{2}}{2\mathcal{H}^{2}},
\end{equation}
and thus (\ref{bgeom_tc}) can be recast into
\begin{equation}
\epsilon=\tilde{\epsilon}-\frac{\mathcal{K}}{2\mathcal{H}^{2}}. \label{epsilon_def}
\end{equation}

Note it is possible to have a spatially flat background solution (i.e. FRW) with non-vanishing $\bar{\phi}^I = \bar{\phi}^I(\vec{x})$.
In this case, (\ref{Rij_fin}) implies $\bar{f}_{ij} \propto \bar{h}_{ij}$. This can be achieved (e.g.) in ``solid inflation'' \cite{Endlich:2012pz}, where three scalar fields are introduced  with $\bar{\phi}^I = x^i \delta_{i}^{I}$ and $f_{IJ} = \delta_{IJ}$ and thus $\bar{f}_{ij} =\bar{h}_{ij}= \delta_{ij}$. 
In this work, however, we are interested in the spatially inhomogeneous backgrounds.
Another difference of our model from solid inflation is that we also introduce an additional scalar field with $\bar{\Phi} = \bar{\Phi}(\tau)$, which explicitly breaks time diffeomorphism.

\subsection{Perturbation theory}

Now let us study the linear perturbations $A$, $B$, $\zeta$, $E$, $\tilde{B}_i$, $F_i$ and $\gamma_{ij}$ around the above inhomogeneous background.
As in the homogeneous and isotropic background, the tensor modes $\gamma_{ij}$ are gauge invariant, and the combination
	\begin{equation}
	\hat{\zeta}:=\zeta-\frac{1}{3}\bar{\nabla}^{2}E-\frac{\mathcal{H}}{\bar{\Phi}'}\delta\Phi,
	\end{equation}
is the gauge invariant curvature perturbation (see Appendix \ref{sec:gauge} for a discussion).
In order the simplify the calculations, we work in the gauge with $E=F_{i}=\delta\Phi=0$, which completely fixes the gauge freedom. 
Note in this gauge, $\zeta$ coincides with the gauge invariant curvature perturbation.

By expanding the action (\ref{model}) around the inhomogeneous background up to the second order in perturbations and solving the non-physical variables, we get the quadratic order action for the perturbations $\zeta$, $\gamma_{ij}$ and $\delta\phi^{I}$:
	\begin{equation}
	S_{2}=\int\mathrm{d}\tau\mathrm{d}^{3}x\sqrt{\bar{h}}\mathcal{L}_{2}=\int\mathrm{d}\tau\mathrm{d}^{3}x\sqrt{\bar{h}}\left(\mathcal{L}_{2}^{(\zeta\zeta)}+\mathcal{L}_{2}^{(\gamma\gamma)}+\mathcal{L}_{2}^{(\delta\phi\delta\phi)}+\mathcal{L}_{2}^{(\zeta\delta\phi)}+\mathcal{L}_{2}^{(\gamma\delta\phi)}\right),\label{L2_fin}
	\end{equation}
where the detailed form for $\mathcal{L}_2$ can be found in (\ref{L2_zz})-(\ref{L2_zf}).
Generally, one needs to solve perturbations on the inhomogeneous background through (\ref{L2_fin}), which is hard to deal with.
On the other hand, we do not expect the observed universe significantly deviates from a FRW one on the large scales. 
What is interesting to us is the leading order corrections to the power spectra of $\zeta$ and $\gamma_{ij}$ due to the background inhomogeneities. 
We thus assume the background inhomogeneities are small and treat the deviation of the background from a homogeneous one as ``perturbation'' as well. 
Supposing the deviation of the inhomogeneous background spatial metric from the flat one is parametrized by
	\begin{equation}
	\bar{h}_{ij}=\delta_{ij}+\kappa X_{ij},\label{hbar_exp}
	\end{equation}
where $X_{ij}$ is time-independent according to our ansatz.
In (\ref{hbar_exp}) we introduce a formal parameter $\kappa$ to characterize the level of background inhomogeneity. 
The perturbative orders of various quantities are
	\begin{equation}
	\bar{\mathcal{R}}_{ij}\sim\bar{f}_{ij}\sim X_{ij}\sim\left|\vec{\partial}\phi^I\right|^{2}\sim\bar{\Gamma}_{ij}^{k}\sim\mathcal{O}\left(\kappa\right).\label{inhomo_order}
	\end{equation}
In this work, we will calculate the corrections to the power spectra of $\zeta$ and $\gamma_{ij}$ up to the linear order in $\kappa$.

We will focus on the effects due to the background inhomogeneities and thus make two assumptions in order to simplify the calculations. First we assume the metric of the field space to be flat, i.e., $f_{IJ}=\delta_{IJ}$. In particular, we have $\bar{f}_{ij}=\delta_{IJ}\partial_{i}\bar{\phi}^{I}\partial_{j}\bar{\phi}^{J}$.
Second we set the constant $\mathcal{K}=0$ in (\ref{bgeom_A_sp}), which also implies $\tilde{\epsilon}=\epsilon$.

We may view our original action as being controlled by two parameters $\delta$ and $\kappa$. The parameter $\delta$ denotes the order of ``quantum fluctuations'' $\zeta$, $\gamma_{ij}$ etc. around the  inhomogeneous background. The other parameter $\kappa$ denotes the deviation of the background from a homogeneous and isotropic one.
Schematically, the original Lagrangian is expanded as
	\begin{equation}
	\mathcal{L}\left(\delta,\kappa\right)= \mathcal{L}_{0}\left(\kappa\right)+ \delta\, \mathcal{L}_{1}\left(\kappa\right)+\delta^2 \mathcal{L}_{2}\left(\kappa\right)+ \mathcal{O}(\delta^3),
	\end{equation}
in which $\mathcal{L}_{2}(\kappa)$ is the quadratic Lagrangian in (\ref{L2_fin}) for quantum fluctuations, but with inhomogeneous coefficients.
Schematically, the quadratic Lagrangian can be written as
\begin{equation}
\mathcal{L}_{2}=\sum_{a,b}\varphi_{a}\hat{\mathcal{O}}_{ab}\varphi_{b},
\end{equation}
where $\varphi_{a}=\left\{ \zeta,\gamma_{ij},\delta\phi^{I}\right\}$, and $\hat{\mathcal{O}}_{ab}$ stands for functions or operators depending on the background quantities, of which the expressions can be read from (\ref{L2_zz})-(\ref{L2_zf}).
In this paper we evaluate the corrections of the inhomogeneous background to the quantum fluctuations up to the leading order in $\kappa$.
To this end, we further expand $\hat{\mathcal{O}}_{ab}$ around its homogeneous part as
	\begin{equation}
		\hat{\mathcal{O}}_{ab}=\hat{\mathcal{O}}_{ab}^{(0)}+\kappa\hat{\mathcal{O}}_{ab}^{(1)}+\mathcal{O}(\kappa^2).
	\end{equation}
Following this strategy, the quadratic order action $S_2$ in (\ref{L2_fin}) can be expanded as
	\begin{eqnarray}
	S_{2}^{(\zeta\zeta)} & = & S_{2,0}^{(\zeta\zeta)}+\kappa S_{2,1}^{(\zeta\zeta)}+\mathcal{O}\left(\kappa^{2}\right),\label{S2_zz_xpd}\\
	S_{2}^{(\gamma\gamma)} & = & S_{2,0}^{(\gamma\gamma)}+\kappa S_{2,1}^{(\gamma\gamma)}+\mathcal{O}\left(\kappa^{2}\right),\label{S2_gg_xpd}\\
	S_{2}^{(\delta\phi\delta\phi)} & = & S_{2,0}^{(\delta\phi\delta\phi)}+\kappa S_{2,1}^{(\delta\phi\delta\phi)}+\mathcal{O}\left(\kappa^{2}\right),\label{S2_ff_xpd}
	\end{eqnarray}
	and
	\begin{eqnarray}
	S_{2}^{(\zeta\delta\phi)} & = & \sqrt{\kappa}\left(S_{2,1}^{(\zeta\delta\phi)}+\mathcal{O}\left(\kappa\right)\right),\label{S2_zf_xpd}\\
	S_{2}^{(\gamma\delta\phi)} & = & \sqrt{\kappa}\left(S_{2,1}^{(\gamma\delta\phi)}+\mathcal{O}\left(\kappa\right)\right).\label{S2_gf_xpd}
	\end{eqnarray}
In the above, various contributions are:
\begin{itemize}
	\item $S_{2,0}^{(\zeta\zeta)}$, $S_{2,0}^{(\gamma\gamma)}$ and $S_{2,0}^{(\delta\phi\delta\phi)}$ are the homogeneous parts of the quadratic order action, which are the same as those in perturbation theory with a FRW background. Precisely, we have
		\begin{equation}
			S_{2,0}^{(\zeta\zeta)}=\int\mathrm{d}\tau\mathrm{d}^{3}x\,\mathcal{L}_{2,0}^{(\zeta\zeta)},\qquad S_{2,0}^{(\gamma\gamma)}=\int\mathrm{d}\tau\mathrm{d}^{3}x\,\mathcal{L}_{2,0}^{(\gamma\gamma)},\qquad S_{2,0}^{(\delta\phi\delta\phi)}=\int\mathrm{d}\tau\mathrm{d}^{3}x\,\mathcal{L}_{2,0}^{(\delta\phi\delta\phi)},
		\end{equation}
	with
		\begin{eqnarray}
		\mathcal{L}_{2,0}^{(\zeta\zeta)} & = & a^{2}\tilde{\epsilon}\left(\zeta'^{2}+\zeta\partial^{2}\zeta\right),\label{S2_zz_0}\\
		\mathcal{L}_{2,0}^{(\gamma\gamma)} & = & \frac{a^{2}}{8}\left(\gamma_{ij}'\gamma_{ij}'+\gamma_{ij}\partial^{2}\gamma_{ij}\right),\label{S2_gg_0}\\
		\mathcal{L}_{2,0}^{(\delta\phi\delta\phi)} & = & \frac{a^{2}}{2}\delta_{IJ}\left(\delta\phi'^{I}\delta\phi'^{J}-\partial_{i}\delta\phi^{I}\partial_{i}\delta\phi^{J}\right),\label{S2_ff_0}
		\end{eqnarray}
		where and in what follows, $\partial^2\equiv \delta^{ij}\partial_{i}\partial_{j}$ and repeated lower spatial indices are summed by $\delta^{ij}$. 
	\item $S_{2,1}^{(\zeta\zeta)}$, $S_{2,1}^{(\gamma\gamma)}$ and $S_{2,1}^{(\delta\phi\delta\phi)}$ can be viewed as ``two-point self-couplings'' of $\zeta$, $\gamma_{ij}$ and $\delta\phi^I$, which arise due to the background inhomogeneities.
	Their couplings are of order $\kappa$.
	For our purpose to calculate the leading order corrections to the power spectra of $\zeta$ and $\gamma_{ij}$, only $S_{2,1}^{(\zeta\zeta)}$ and $S_{2,1}^{(\gamma\gamma)}$ are needed, which are given by
		\begin{equation}
			S_{2,1}^{(\zeta\zeta)}=\int\mathrm{d}\tau\mathrm{d}^{3}x\,\mathcal{L}_{2,1}^{(\zeta\zeta)},\qquad S_{2,1}^{(\gamma\gamma)}=\int\mathrm{d}\tau\mathrm{d}^{3}x\,\mathcal{L}_{2,1}^{(\gamma\gamma)},
		\end{equation}
	with
		\begin{equation}
		\mathcal{L}_{2,1}^{(\zeta\zeta)}=\frac{1}{2}a^{2}\epsilon\left(X_{ii}\zeta'^{2}-X_{jj}\,\partial_{i}\zeta\partial_{i}\zeta+2X_{ij}\,\partial_{i}\zeta\partial_{j}\zeta\right),\label{hL2zz_1}
		\end{equation}
	and
		\begin{eqnarray}
		\mathcal{L}_{2,1}^{(\gamma\gamma)} & = & \frac{a^{2}}{16}\Big[X_{kk}\gamma_{ij}'\gamma_{ij}'-4X_{ij}\gamma_{ik}'\gamma_{jk}'-X_{kk}\partial_{l}\gamma_{ij}\partial_{l}\gamma_{ij}+4X_{ik}\partial_{l}\gamma_{kj}\partial_{l}\gamma_{ij}+2X_{kl}\partial_{k}\gamma_{ij}\partial_{l}\gamma_{ij}\nonumber \\
		&  & -4\left(\partial_{i}X_{lk}-\partial_{l}X_{ki}\right)\gamma_{ij}\partial_{k}\gamma_{lj}-2\left(6\partial_{k}\partial_{i}X_{jk}-2\partial^{2}X_{ij}-3\partial_{i}\partial_{j}X_{kk}\right)\gamma_{il}\gamma_{lj}\nonumber \\
		&  & +2\left(\partial_{k}\partial_{l}X_{kl}-\partial^{2}X_{kk}+2\mathcal{K}\right)\gamma_{ij}\gamma_{ij}\Big].\label{hL2gg_1}
		\end{eqnarray}
	\item $S_{2,1}^{(\zeta\delta\phi)}$ and $S_{2,1}^{(\gamma\delta\phi)}$ are ``two-point cross-interactions'' between $\zeta$, $\gamma_{ij}$ and $\delta\phi$, which are given by
		\begin{eqnarray}
		\mathcal{L}_{2,1}^{(\zeta\delta\phi)} & = & -a^{2}\epsilon\left[\left(\partial_{i}\partial^{-2}\zeta'\right)\left(\delta_{IJ}\partial_{i}\bar{\phi}^{J}\delta\phi'^{I}\right)+\partial_{i}\zeta\left(\delta_{IJ}\partial_{i}\bar{\phi}^{J}\delta\phi^{I}\right)\right],\label{hL2zf_1}\\
		\mathcal{L}_{2,1}^{(\gamma\delta\phi)} & = & -a^{2}\gamma_{ij}\delta_{IJ}\partial_{i}\partial_{j}\bar{\phi}^{J}\delta\phi^{I}.\label{hL2gf_1}
		\end{eqnarray}
	Due to the inhomogeneous nature of the background geometry, there arises the mixing between the scalar modes $\delta\phi^I$ and the tensor modes $\gamma_{ij}$. 
\end{itemize}

In the following, we treat  $\mathcal{L}_{2,1}^{(\zeta\zeta)}$, $\mathcal{L}_{2,1}^{(\gamma\gamma)}$, $\mathcal{L}_{2,1}^{(\zeta\delta\phi)}$ and $\mathcal{L}_{2,1}^{(\gamma\delta\phi)}$ as ``two-point interactions'' and employ the in-in formalism to calculate the  leading order corrections to the power spectra of $\zeta$ and $\gamma_{ij}$.

Up to the linear order in $\kappa$, the conjugate momentum for $\gamma_{ij}$ is given by
	\begin{equation}
	\pi_{ij}\supset \frac{\delta}{\delta\gamma_{ij}'}\int\mathrm{d}^{3}x\left(\mathcal{L}_{2,0}^{(\gamma\gamma)}+\mathcal{L}_{2,1}^{(\gamma\gamma)} + \mathcal{L}_{2,1}^{(\gamma\delta\phi)}\right)=\frac{a^{2}}{4}\gamma_{ij}'+\frac{a^{2}}{8}X_{kk}\gamma_{ij}'-\frac{a^{2}}{2}X_{k(i}\gamma_{j)k}',\label{pi_ij}
	\end{equation}
where we used $\frac{\partial\gamma_{ij}}{\partial\gamma_{kl}}=\frac{1}{2}\left(\delta_{ik}\delta_{jl}+\delta_{il}\delta_{jk}\right)$.
Similarly, the conjugate momenta for $\zeta$ and $\delta\phi^{I}$ are given by 
	\begin{equation}
	\pi_{\zeta}  \supset  \frac{\delta}{\delta\zeta'}\int\mathrm{d}^{3}x\left(\mathcal{L}_{2,0}^{(\zeta\zeta)}+\mathcal{L}_{2,1}^{(\zeta\zeta)}+\mathcal{L}_{2,1}^{(\zeta\delta\phi)}\right) 
	 \simeq  a^{2}\epsilon\left[2\zeta'+X_{ii}\zeta'+\partial^{-2}\partial_{i}\left(\partial_{i}\bar{\phi}^{I}\delta\phi'^{I}\right)\right],\label{pi_zeta}
	\end{equation}
and
	\begin{equation}
		\pi^{I}\supset\frac{\delta}{\delta\phi'^{I}}\int\mathrm{d}^{3}x\left(\mathcal{L}_{2,0}^{(\delta\phi\delta\phi)}+\mathcal{L}_{2,1}^{(\zeta\delta\phi)}\right)=a^{2}\left[\delta\phi'^{I}-\tilde{\epsilon}\,\partial_{i}\bar{\phi}^{I}\left(\partial_{i}\partial^{-2}\zeta'\right)\right], \label{pi^I}
	\end{equation}
respectively.
In order to calculate the Hamiltonian, we need to solve the ``velocities'' $\gamma_{ij}'$, $\zeta'$ and $\delta'^{I}$ in terms of their conjugate
momenta by reverting (\ref{pi_ij}), (\ref{pi_zeta}) and (\ref{pi^I}). 
To this end, we write
	\begin{equation}
		\gamma_{ij}'=\left(\gamma_{ij}'\right)_{(0)}+\kappa\left(\gamma_{ij}'\right)_{\left(1\right)}+\mathcal{O}\left(\kappa^{2}\right), \label{gmp_inv}
	\end{equation}
where subscript ``$_{\left(i\right)}$'' denotes the order in $\kappa$.
Plugging (\ref{gmp_inv}) into (\ref{pi_ij}), up to the linear order in $\kappa$ we get
	\begin{equation}
	\gamma_{ij}' \supset \frac{1}{a^{2}}\left(4\pi_{ij}-2X_{kk}\pi_{ij}+8X_{k(i}\pi_{j)k}\right).\label{gamma_ijp_pi}
	\end{equation}
Similarly,
	\begin{eqnarray}
	\zeta' & \supset & \frac{1}{2a^{2}\epsilon}\pi_{\zeta}-\frac{1}{4a^{2}\epsilon}X_{ii}\pi_{\zeta}-\frac{1}{2a^{2}}\,\partial^{-2}\partial_{i}\left(\partial_{i}\bar{\phi}^{I}\pi^{I}\right),\label{zetap_pi}\\
	\delta\phi'^{I} & \supset & \frac{1}{a^{2}}\pi^{I}+\frac{1}{2a^{2}}\partial_{i}\bar{\phi}^{I}\left(\partial_{i}\partial^{-2}\pi_{\zeta}\right).\label{dfp_pi}
	\end{eqnarray}

We can straightly find the corresponding Hamiltonian in the interaction picture as 
	\begin{equation}
	H=\int\mathrm{d}^{3}x\left(\pi_{ij}\gamma_{ij}'+\pi_{\zeta}\zeta'+\pi^{I}\delta\phi'^{I}-\sqrt{\bar{h}}\mathcal{L}_{2}\right),\label{Ham_def}
	\end{equation}
where $\mathcal{L}_2$ is given in (\ref{L2_fin}).
Using the above results, up to the linear order in $\kappa$, the Hamiltonian is given by (in momentum space)
	\begin{equation}
	H=H_0+H_1,
	\end{equation}
where we have split the Hamiltonian into a ``free'' part $H_0$ and a ``two-point interaction'' part $H_1$.
The free Hamiltonian $H_0$ is given by
	\begin{equation}
		H_0 = \int \mathrm{d}^3x\,\left(\mathcal{H}_{0}^{\left(\gamma\gamma\right)}+\mathcal{H}_{0}^{\left(\zeta\zeta\right)}+\mathcal{H}_{0}^{\left(\delta\phi\delta\phi\right)}\right),
	\end{equation}
with
	\begin{eqnarray}
	\mathcal{H}_{0}^{\left(\gamma\gamma\right)} & = & \frac{2}{a^{2}}\pi_{ij}^{2}-\frac{a^{2}}{8}\gamma_{ij}\partial^{2}\gamma_{ij},\label{Ham_gg_0}\\
	\mathcal{H}_{0}^{\left(\zeta\zeta\right)} & = & \frac{1}{4a^{2}\epsilon}\pi_{\zeta}^{2}-a^{2}\epsilon\,\zeta\,\partial^{2}\zeta,\label{Ham_zz_0}\\
	\mathcal{H}_{0}^{\left(\delta\phi\delta\phi\right)} & = & \frac{1}{2a^{2}}\pi^{I}\pi^{I}+\frac{a^{2}}{2}\partial_{i}\delta\phi^{I}\partial_{i}\delta\phi^{I},\label{Ham_ff_0}
	\end{eqnarray}
which are exactly the same as in the case of homogeneous inflation with multiple scalar fields.
For our purpose to calculate the corrections to the power spectra of $\gamma_{ij}$ and $\zeta$, the relevant contributions to the ``two-point interaction'' Hamiltonian $H_1$ are 
\begin{equation}
H_{1} \supset  H_{1}^{\left(\gamma\gamma\right)}+H_{1}^{\left(\zeta\zeta\right)}+H_{1}^{\left(\zeta\delta\phi\right)}+H_{1}^{\left(\gamma\delta\phi\right)}.
\end{equation}
where the first two terms are the self-couplings of $\gamma_{ij}$ and $\zeta$, and the last two terms are the cross-couplings of $\gamma_{ij}$ and $\zeta$ with the scalar field perturbations $\delta\phi^{I}$ due to the background inhomogeneities.
Straightforward calculations show that
	\begin{eqnarray}
	\mathcal{H}_{1}^{\left(\gamma\gamma\right)} & = & -\frac{1}{a^{2}}X_{kk}\pi_{ij}^{2}+\frac{4}{a^{2}}X_{ij}\pi_{ik}\pi_{jk}+\frac{a^{2}}{16}X_{kk}\left(\partial_{l}\gamma_{ij}\right)^{2}-\frac{a^{2}}{4}X_{ik}\partial_{l}\gamma_{kj}\partial_{l}\gamma_{ij}-\frac{a^{2}}{8}X_{kl}\partial_{k}\gamma_{ij}\partial_{l}\gamma_{ij}\nonumber \\
	&  & +\frac{a^{2}}{4}\left(\partial_{i}X_{lk}-\partial_{l}X_{ki}\right)\gamma_{ij}\partial_{k}\gamma_{lj}+\frac{a^{2}}{8}\left(6\partial_{k}\partial_{i}X_{jk}-2\partial^{2}X_{ij}-3\partial_{i}\partial_{j}X_{kk}\right)\gamma_{il}\gamma_{lj}\nonumber \\
	&  & -\frac{a^{2}}{8}\left(\partial_{k}\partial_{l}X_{kl}-\partial^{2}X_{kk}\right)\gamma_{ij}^{2},\label{H1_gg}
	\end{eqnarray}
	\begin{equation}
	\mathcal{H}_{1}^{\left(\zeta\zeta\right)}=-\frac{1}{8a^{2}\epsilon}X_{ii}\pi_{\zeta}^{2}+\frac{1}{8a^{2}}\bar{f}_{ij}\left(\partial_{i}\partial^{-2}\pi_{\zeta}\right)\left(\partial_{j}\partial^{-2}\pi_{\zeta}\right)+a^{2}\epsilon\left(\frac{1}{2}X_{jj}\,\partial_{i}\zeta\partial_{i}\zeta-X_{ij}\,\partial_{i}\zeta\partial_{j}\zeta\right),\label{H1_zz}
	\end{equation}
	\begin{equation}
	\mathcal{H}_{1}^{\left(\zeta\delta\phi\right)}=\frac{1}{2a^{2}}\partial_{i}\bar{\phi}^{I}\pi^{I}\left(\partial_{i}\partial^{-2}\pi_{\zeta}\right)+a^{2}\epsilon\,\partial_{i}\zeta\left(\partial_{i}\bar{\phi}^{I}\delta\phi^{I}\right),\label{H1_zf}
	\end{equation}
	\begin{equation}
	\mathcal{H}_{1}^{\left(\gamma\delta\phi\right)}=a^{2}\partial_{i}\partial_{j}\bar{\phi}^{I}\,\delta\phi^{I}\,\gamma_{ij}.\label{H1_gf}
	\end{equation}

In order to canonically quantize the system, for the tensor perturbations, we write
	\begin{equation}
		\hat{\gamma}_{ij}\left(\tau,\bm{k}\right) = \frac{2}{a\left(\tau\right)}\sum_{s=\pm2}\left(u\left(\tau,\bm{k}\right)e_{ij}^{\left(s\right)}(\hat{\bm{k}})\hat{a}_{s}\left(\bm{k}\right)+u^{\ast}\left(\tau,-\bm{k}\right)e_{ij}^{\left(s\right)\ast}(-\hat{\bm{k}})\hat{a}_{s}^{\dagger}\left(-\bm{k}\right)\right),\label{gamma_dec}
	\end{equation}
where an asterisk denotes complex conjugate, $\hat{a}_{s}\left(\bm{k}\right)$ and $\hat{a}_{s}^{\dagger}\left(\bm{k}\right)$ are the annihilation and creation operators for tensor perturbations with the commutation relation
	\begin{equation}
	\left[\hat{a}_{s}\left(\bm{k}\right),\hat{a}_{s'}^{\dagger}\left(\bm{k}'\right)\right]=\left(2\pi\right)^{3}\delta_{ss'}\delta^{3}\left(\bm{k}-\bm{k}'\right),\label{cc_t}
	\end{equation}
$e_{ij}^{\left(s\right)}(\hat{\bm{k}})$ is the polarization tensor with the helicity states $s = \pm 2$, satisfying
	\begin{equation}
	\sum_{i}e_{ii}^{\left(s\right)}(\hat{\bm{k}})=\sum_{i}k_{i}e_{ij}^{\left(s\right)}(\hat{\bm{k}})=0,\qquad e_{ij}^{\left(s\right)\ast}(\hat{\bm{k}})=e_{ij}^{\left(-s\right)}(\hat{\bm{k}})=e_{ij}^{\left(s\right)}(-\hat{\bm{k}}).\label{eij}
	\end{equation}
By choosing the normalization condition
	\begin{equation}
	\sum_{i,j}e_{ij}^{\left(s\right)}(\hat{\bm{k}})e_{ij}^{\left(s'\right)\ast}(\hat{\bm{k}})=\delta^{ss'},\label{eijnom}
	\end{equation}
$\frac{a}{\sqrt{2}}\hat{\gamma}_{ij}$ is canonically normalized, of which the mode function $u(\tau,\bm{k})$ is given by
	\begin{equation}
	u\left(\tau,\bm{k}\right)=u\left(\tau,k\right)=\frac{1}{\sqrt{2k}}\left(1-\frac{i}{k\tau}\right)e^{-ik\tau},\label{ms_u}
	\end{equation}
where we have used the de Sitter approximation $a\left(\tau\right)=-1/(H\tau)$.

The canonical quantization of $\zeta$ and $\delta\phi^I$ is completely parallel.
We write
	\begin{eqnarray}
	\hat{\zeta}\left(\tau,\bm{k}\right) & = & \frac{1}{a\left(\tau\right)\sqrt{2\epsilon}}\left(v\left(\tau,\bm{k}\right)\hat{b}\left(\bm{k}\right)+v^{\ast}\left(\tau,-\bm{k}\right)\hat{b}^{\dagger}\left(-\bm{k}\right)\right),\label{zeta_dec}\\
	\delta\hat{\phi}^{I}\left(\tau,\bm{k}\right) & = & \frac{1}{a\left(\tau\right)}\left(w\left(\tau,\bm{k}\right)\hat{c}^{I}\left(\bm{k}\right)+w^{\ast}\left(\tau,-\bm{k}\right)\hat{c}^{I\dagger}\left(-\bm{k}\right)\right),\label{phi_dec}
	\end{eqnarray}
where $\hat{b}\left(\bm{k}\right)$, $\hat{c}^{I}\left(\bm{k}\right)$ and $\hat{b}^{\dagger}\left(\bm{k}\right)$, $\hat{c}^{I\dagger}\left(\bm{k}\right)$ are the  annihilation and creation operators for $\hat{\zeta}$ and $\delta\hat{\phi}^{I}$, satisfying
	\begin{equation}
	\left[\hat{b}\left(\bm{k}\right),\hat{b}^{\dagger}\left(\bm{k}'\right)\right]=\left(2\pi\right)^{3}\delta^{3}\left(\bm{k}-\bm{k}'\right),\qquad\left[\hat{c}^{I}\left(\bm{k}\right),\hat{c}^{J\dagger}\left(\bm{k}'\right)\right]=\left(2\pi\right)^{3}\delta^{IJ}\delta^{3}\left(\bm{k}-\bm{k}'\right),\label{bc_comm}
	\end{equation}
and
	\begin{equation}
	v\left(\tau,\bm{k}\right)=w\left(\tau,\bm{k}\right)=\frac{1}{\sqrt{2k}}\left(1-\frac{i}{k\tau}\right)e^{-ik\tau}.\label{ms_vw}
	\end{equation}

Before going into the concrete calculations, note terms in $H_1$ take the following general structure in momentum space
	\begin{equation}
	H_{1}^{(\gamma\gamma)}=\int\widetilde{\mathrm{d}\bm{k}_{123}}\left(\frac{1}{a^{2}}\pi_{ij}(\bm{k}_{1})\pi_{kl}(\bm{k}_{2})\mathcal{C}_{ij,kl}(\bm{k}_{1},\bm{k}_{2},\bm{k}_{3})+a^{2}\gamma_{ij}(\bm{k}_{1})\gamma_{kl}(\bm{k}_{2})\mathcal{D}_{ji,kl}(\bm{k}_{1},\bm{k}_{2},\bm{k}_{3})\right),\label{Ham_gg_1}
	\end{equation}
	\begin{equation}
	H_{1}^{\left(\zeta\zeta\right)}=\int\widetilde{\mathrm{d}\bm{k}_{123}}\left(\frac{1}{a^{2}\epsilon}\pi_{\zeta}(\bm{k}_{1})\pi_{\zeta}(\bm{k}_{2})\mathcal{C}(\bm{k}_{1},\bm{k}_{2},\bm{k}_{3})+a^{2}\epsilon\,\zeta(\bm{k}_{1})\zeta(\bm{k}_{2})\mathcal{D}(\bm{k}_{1},\bm{k}_{2},\bm{k}_{3})\right),\label{Ham_zz_1}
	\end{equation}
	\begin{equation}
	H_{1}^{\left(\zeta\delta\phi\right)}=\int\widetilde{\mathrm{d}\bm{k}_{123}}\left(\frac{1}{a^{2}}\pi_{\zeta}(\bm{k}_{1})\pi^{I}(\bm{k}_{2})\mathcal{C}^{I}(\bm{k}_{1},\bm{k}_{2},\bm{k}_{3})+a^{2}\epsilon\,\zeta(\bm{k}_{1})\delta\phi^{I}(\bm{k}_{2})\mathcal{D}^{I}(\bm{k}_{1},\bm{k}_{2},\bm{k}_{3})\right),\label{Ham_zf_1}
	\end{equation}
	\begin{equation}
	H_{1}^{\left(\gamma\delta\phi\right)}=\int\widetilde{\mathrm{d}\bm{k}_{123}}\,a^{2}\gamma_{ij}(\bm{k}_{1})\delta\phi^{I}(\bm{k}_{2})\mathcal{D}_{ij}^{I}(\bm{k}_{1},\bm{k}_{2},\bm{k}_{3}),\label{Ham_gf_1}
	\end{equation}
where $\widetilde{\mathrm{d}\bm{k}_{123}}$ is a shorthand for
	$\frac{\mathrm{d}^{3}k_{1}}{\left(2\pi\right)^{3}}\frac{\mathrm{d}^{3}k_{2}}{\left(2\pi\right)^{3}}\frac{\mathrm{d}^{3}k_{3}}{\left(2\pi\right)^{3}}\left(2\pi\right)^{3}\delta^{3}\left(\bm{k}_{1}+\bm{k}_{2}+\bm{k}_{3}\right)$.
Variously defined $\mathcal{C}$'s and $\mathcal{D}$'s in (\ref{Ham_gg_1})-(\ref{Ham_gf_1}) are approximately time-independent, which we assume to be fully symmetrized. 
That is, all $\mathcal{C}$'s and $\mathcal{D}$'s are symmetric with respect to $\bm{k}_{1}\leftrightarrow \bm{k}_{2}$, and 
 $\mathcal{C}_{ij,kl}$ and $\mathcal{D}_{ij,kl}$ are symmetric under  exchanging of indices
	\begin{eqnarray}
	\mathcal{C}_{ij,kl} & = & \mathcal{C}_{ji,kl}=\mathcal{C}_{ij,lk}=\mathcal{C}_{kl,ij},\label{calCijkl_sym}\\
	\mathcal{D}_{ij,kl} & = & \mathcal{D}_{ji,kl}=\mathcal{D}_{ij,lk}=\mathcal{D}_{kl,ij},\label{calDijkl_sym}
	\end{eqnarray}
as well.

In our model, the $\mathcal{C}$'s and $\mathcal{D}$'s factors can be read from (\ref{H1_gg})-(\ref{H1_gf}) after going into the momentum space. Precisely,
	\begin{eqnarray}
	\mathcal{C}_{ij,kl}\left(\bm{k}_{1},\bm{k}_{2},\bm{k}_{3}\right) & = & -\frac{1}{2}(\delta_{ik}\delta_{jl}+\delta_{il}\delta_{jk})X_{mm}(\bm{k}_{3})\nonumber\\
	&  & +\delta_{jk}X_{il}(\bm{k}_{3})+\delta_{ik}X_{jl}(\bm{k}_{3})+\delta_{li}X_{kj}(\bm{k}_{3})+\delta_{lj}X_{ki}(\bm{k}_{3}),\label{calC_ijkl}
	\end{eqnarray}
and
	\begin{eqnarray}
	&  & \mathcal{D}_{ij,kl}\left(\bm{k}_{1},\bm{k}_{2},\bm{k}_{3}\right)\nonumber \\
	& = & \frac{1}{64}\Big[-\frac{1}{4}\left(\bm{k}_{1}\cdot\bm{k}_{2}\right)\delta_{ik}\delta_{jl}X_{mm}\left(\bm{k}_{3}\right)+\left(\bm{k}_{1}\cdot\bm{k}_{2}\right)\delta_{jk}X_{li}\left(\bm{k}_{3}\right)\nonumber \\
	&  & \quad+\frac{1}{2}\delta_{ik}\delta_{jl}\left(\bm{k}_{1m}\bm{k}_{2n}\right)X_{mn}\left(\bm{k}_{3}\right)+\delta_{jk}\left(-\bm{k}_{2m}\bm{k}_{3i}X_{lm}\left(\bm{k}_{3}\right)+\bm{k}_{2m}\bm{k}_{3l}X_{mi}\left(\bm{k}_{3}\right)\right)\nonumber \\
	&  & \quad+\frac{1}{2}\delta_{jl}\left(-6\bm{k}_{3m}\bm{k}_{3i}X_{km}\left(\bm{k}_{3}\right)+2k_{3}^{2}X_{ik}\left(\bm{k}_{3}\right)+3\bm{k}_{3i}\bm{k}_{3k}X_{mm}\left(\bm{k}_{3}\right)\right)\nonumber \\
	&  & \quad+\frac{1}{2}\delta_{ik}\delta_{jl}\left(\bm{k}_{3m}\bm{k}_{3n}X_{mn}\left(\bm{k}_{3}\right)-k_{3}^{2}X_{mm}\left(\bm{k}_{3}\right)\right).\nonumber \\
	&  & \quad+\text{symm. with }\left\{ _{i}\leftrightarrow{}_{j}\right\} +\left\{ _{k}\leftrightarrow{}_{l}\right\} +\left\{ _{i,j}\leftrightarrow{}_{k,l}\right\} +\left\{ \bm{k}_{1}\leftrightarrow\bm{k}_{2}\right\} \Big],\label{calD_ijkl}
	\end{eqnarray}
and
	\begin{eqnarray}
	\mathcal{C}\left(\bm{k}_{1},\bm{k}_{2},\bm{k}_{3}\right) & = & -\frac{1}{8}X_{ii}\left(\bm{k}_{3}\right)-\frac{\epsilon}{8}\frac{1}{k_{1}^{2}}\frac{1}{k_{2}^{2}}\bm{k}_{1i}\bm{k}_{2j}\bar{f}_{ij}\left(\bm{k}_{3}\right),\label{calC_fin}\\
	\mathcal{D}\left(\bm{k}_{1},\bm{k}_{2},\bm{k}_{3}\right) & = & -\frac{1}{2}\left(\bm{k}_{1}\cdot\bm{k}_{2}\right)X_{jj}\left(\bm{k}_{3}\right)+\left(\bm{k}_{1i}\bm{k}_{2j}\right)X_{ij}\left(\bm{k}_{3}\right).\label{calD_fin}
	\end{eqnarray}
and
	\begin{equation}
	\mathcal{C}^{I}\left(\bm{k}_{1},\bm{k}_{2},\bm{k}_{3}\right)=\frac{1}{2}\frac{\bm{k}_{1}\cdot\bm{k}_{3}}{k_{1}^{2}}\bar{\phi}^{I}\left(\bm{k}_{3}\right),\qquad\mathcal{D}^{I}\left(\bm{k}_{1},\bm{k}_{2},\bm{k}_{3}\right)=-\bm{k}_{1}\cdot\bm{k}_{3}\bar{\phi}^{I}\left(\bm{k}_{3}\right),\label{calCIDI_fin}
	\end{equation}
	\begin{equation}
	\mathcal{D}_{ij}^{I}\left(\bm{k}_{1},\bm{k}_{2},\bm{k}_{3}\right)=-k_{3i}k_{3j}\bar{\phi}^{I}\left(\bm{k}_{3}\right).\label{calDIij_fin}
	\end{equation}

In this work, we focus on background configurations in which $\mathcal{C}$'s and $\mathcal{D}$ are real functions. 
According to the above, $X_{ij}(\bm{k})$ and $\bar{\phi}^{I}(\bm{k})$ must satisfy
	\begin{equation}
	X_{ij}^{\ast}\left(\bm{k}\right)=X_{ij}\left(-\bm{k}\right)=X_{ij}\left(\bm{k}\right),\qquad\bar{\phi}^{I\ast}\left(\bm{k}\right)=\bar{\phi}^{I}\left(-\bm{k}\right)=\bar{\phi}^{I}\left(\bm{k}\right). \label{Xphi_rc}
	\end{equation}

\section{Gravitational waves} \label{sec:gw}

In this section, we calculate the correction to the power spectra of tensor perturbations due to the background inhomogeneities.
Using (\ref{gamma_dec}),  the ``free''  two-point function  for the tensor perturbations is
	\begin{equation}
	\left\langle \hat{\gamma}_{ij}\left(\tau,\bm{k}\right)\hat{\gamma}_{kl}\left(\tau',\bm{k}'\right)\right\rangle^{(0)} =\left(2\pi\right)^{3}\delta^{3}\left(\bm{k}+\bm{k}'\right) 4\Pi_{ij,kl}(\hat{\bm{k}})\frac{u\left(\tau,k\right)}{a\left(\tau\right)}\frac{u^{\ast}\left(\tau',k\right)}{a\left(\tau'\right)},\label{2pf_gg}
	\end{equation}
where the mode function $u(\tau,k)$ is given in (\ref{ms_u}) and we denote 
	\begin{equation}
	\Pi_{ij,kl}(\hat{\bm{k}})\equiv\sum_{s=\pm2}e_{ij}^{\left(s\right)}(\hat{\bm{k}})e_{kl}^{\left(s\right)\ast}(\hat{\bm{k}}),\label{Piij}
	\end{equation}
for short. 
It is easy to verify that $\Pi_{ij,kl}(\hat{\bm{k}})$ is real, i.e., $\Pi_{ij,kl}^{\ast}(\hat{\bm{k}})=\Pi_{ij,kl}(\hat{\bm{k}})$, which has the  following properties:
	\begin{equation}
	\Pi_{kl,ij}(\hat{\bm{k}})=\Pi_{ij,kl}(\hat{\bm{k}}),\qquad \sum_{i,j}e_{ij}^{\left(s\right)\ast}(\hat{\bm{k}})\Pi_{ij,kl}(\hat{\bm{k}})=e_{kl}^{\left(s\right)\ast}(\hat{\bm{k}}).\label{Piij_sym}
	\end{equation}

The standard isotropic and homogeneous power spectrum for the tensor modes $\gamma_{ij}$ is defined by
	\begin{equation}
	\Delta_{ij,kl}^{(0)}(\bm{k},\bm{k}') \equiv \left\langle \hat{\gamma}_{ij}(\bm{k})\hat{\gamma}_{kl}(\bm{k}')\right\rangle^{(0)} =\left(2\pi\right)^{3}\delta^{3}(\bm{k}+\bm{k}')\mathcal{P}_{ij,kl}^{(0)}(\bm{k}), \label{Delta_gamma_0}
	\end{equation}
with
	\begin{equation}
	\mathcal{P}_{ij,kl}^{(0)}(\bm{k})=4\,\Pi_{ij,kl}(\hat{\bm{k}})\left.\frac{\left|u\left(\tau,k\right)\right|^{2}}{a^{2}(\tau)}\right|_{\tau\rightarrow0}.
	\end{equation}
Accordingly, the total power spectrum is thus
	\begin{equation}
	\mathcal{P}_{\mathrm{T}}^{(0)}=\frac{k^{3}}{2\pi^{2}}\sum_{i,j}\mathcal{P}_{ij,ij}^{(0)}=\frac{k^{3}}{\pi^{2}}\frac{4}{a^{2}}\left|u\left(\tau,k\right)\right|^{2}.
	\end{equation}

\begin{figure}[h]
		\begin{centering}
			\includegraphics[scale=0.6]{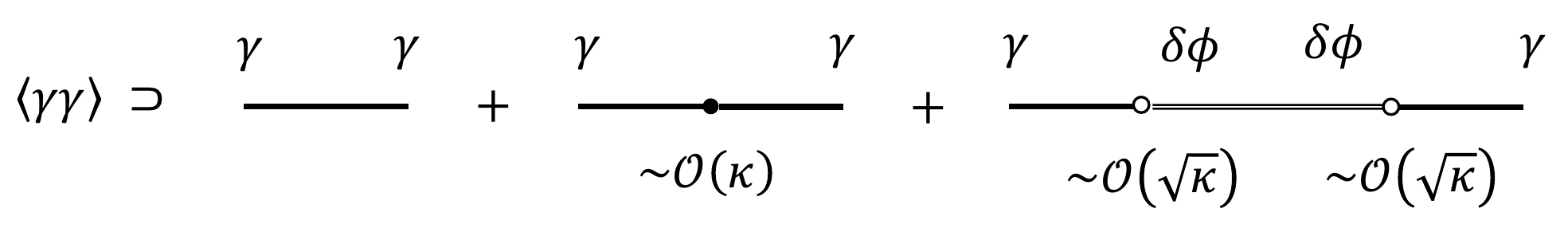}
			\par\end{centering}
		\caption{Illustration of contributions to the power spectra of the tensor perturbations $\gamma_{ij}$ up to the linear order in $\kappa$. The total power spectra consist of three contributions: the standard isotropic and homogeneous spectra (left), corrections due to the inhomogeneous background metric $\bar{h}_{ij}$ (middle), corrections due to the couplings with scalar field perturbations $\delta\phi^{I}$ (right). Both corrections are of order $\kappa$.}
		\label{fig:gamma}
	\end{figure}

According to the standard in-in formalism, up to the linear order in $\kappa$, the  power spectrum of $\gamma_{ij}$ is
	\begin{equation}
		\left\langle \hat{\gamma}_{ij}(\bm{k})\hat{\gamma}_{kl}(\bm{k}')\right\rangle =\left[\Delta_{ij,kl}^{(0)}+\Delta_{ij,kl}^{(1)}+\Delta_{ij,kl}^{(2)}\right](\bm{k},\bm{k}'),
	\end{equation}
where $\Delta_{ij,kl}^{(0)}(\bm{k},\bm{k}')$ is the standard isotropic and homogeneous  power spectrum given in (\ref{Delta_gamma_0}), $\Delta_{ij,kl}^{(1)}(\bm{k},\bm{k}')$ and $\Delta_{ij,kl}^{(2)}(\bm{k},\bm{k}')$ are corrections due to the background inhomogeneities, which are given by
	\begin{equation}
	\Delta_{ij,kl}^{(1)}(\bm{k},\bm{k}')=2\Im\left(\int_{-\infty}^{0}\!\mathrm{d}\tau\big<\hat{\gamma}_{ij}(0,\bm{k})\hat{\gamma}_{kl}(0,\bm{k}')H_{1}^{(\gamma\gamma)}(\tau)\big>\right),\label{inin_gg_1}
	\end{equation}
and
	\begin{eqnarray}
	\Delta_{ij,kl}^{(2)}(\bm{k},\bm{k}') & = & \int_{-\infty}^{0}\!\mathrm{d}\tau_{1}\int_{-\infty}^{0}\!\mathrm{d}\tau_{2}\big<H_{1}^{\left(\gamma\delta\phi\right)}(\tau_{1})\hat{\gamma}_{ij}(0,\bm{k})\hat{\gamma}_{kl}(0,\bm{k}')H_{1}^{\left(\gamma\delta\phi\right)}(\tau_{2})\big>\nonumber \\
	&  & -2\Re\left(\int_{-\infty}^{0}\!\mathrm{d}\tau_{1}\int_{-\infty}^{\tau_{1}}\!\mathrm{d}\tau_{2}\big<\hat{\gamma}_{ij}(0,\bm{k})\hat{\gamma}_{kl}(0,\bm{k}')H_{1}^{(\gamma\delta\phi)}(\tau_{1})H_{1}^{(\gamma\delta\phi)}(\tau_{2})\big>\right),\label{inin_gg_2}
	\end{eqnarray}
where $H_{1}^{\left(\gamma\gamma\right)}$ and $H_{1}^{\left(\gamma\delta\phi\right)}$ are given in (\ref{Ham_gg_1}) and (\ref{Ham_gf_1}), respectively.
These two types of corrections are illustrated in the Feynman-type diagrams in Fig.\ref{fig:gamma}. 
In particular, 
\begin{itemize}
	\item $\Delta_{ij,kl}^{(1)}(\bm{k},\bm{k}')$ is the correction from the inhomogeneous background metric $\bar{h}_{ij} = \delta_{ij}+X_{ij}$ directly,
	\item $\Delta_{ij,kl}^{(2)}(\bm{k},\bm{k}')$ is the correction from the couplings between the tensor perturbations $\gamma_{ij}$ and perturbations of the scalar fields $\delta\phi^{I}$, of which the existence is due to the background inhomogeneity.
\end{itemize}

Plugging the general form for the Hamiltonians (\ref{Ham_gg_1}) and (\ref{Ham_gf_1}) into (\ref{inin_gg_1}) and (\ref{inin_gg_2}), we find
	\begin{eqnarray}
	\Delta_{ij,kl}^{(1)}(\bm{k},\bm{k}') & = & -\Pi_{ij,i'j'}(\hat{\bm{k}})\Pi_{kl,k'l'}(\hat{\bm{k}}')\frac{H^{2}}{k'k(k'+k)}\times\bigg[\mathcal{C}_{i'j',k'l'}(-\bm{k},-\bm{k}',\bm{k}+\bm{k}')\nonumber \\
	&  & +16\,\mathcal{D}_{j'i',k'l'}(-\bm{k},-\bm{k}',\bm{k}+\bm{k}')\frac{k'^{2}+k'k+k^{2}}{k'^{2}k^{2}}\bigg], \label{Delta_ijkl_1}
	\end{eqnarray}
and
	\begin{eqnarray}
	\Delta_{ij,kl}^{(2)}(\bm{k},\bm{k}') & = & 8H^{2}\Pi_{ij,i'j'}(\hat{\bm{k}})\Pi_{kl,k'l'}(\hat{\bm{k}}')\nonumber \\
	&  & \times\int\!\frac{\mathrm{d}^{3}p}{\left(2\pi\right)^{3}}\mathcal{F}(k,k',p)\mathcal{D}_{i'j'}^{I}(-\bm{k},-\bm{p},\bm{k}+\bm{p})\mathcal{D}_{k'l'}^{I}(-\bm{k}',\bm{p},\bm{k}'-\bm{p}), \label{Delta_ijkl_2}
	\end{eqnarray}
where we define
	\begin{eqnarray}
	\mathcal{F}(k,k',p) & = & \frac{1}{k'^{3}k^{3}p^{3}(k'+k)(k'+p)(k+p)}\big[p^{3}(k'^{2}+k'k+k^{2})\nonumber \\
	&  & +k'^{2}k^{2}(k'+k)+p^{2}\left(k'^{3}+2k'^{2}k+2k'k^{2}+k^{3}\right)+k'kp(k'+k)^{2}\big].\label{calF}
	\end{eqnarray}
In deriving the above, we used the assumption that all $\mathcal{C}$'s and $\mathcal{D}$'s factors defined in  (\ref{Ham_gg_1})-(\ref{Ham_gf_1}) are real, and used de Sitter approximation for the homogeneous metric in evaluating the time integrals.  $\Delta_{ij,kl}^{(1)}(\bm{k},\bm{k}')$ and $\Delta_{ij,kl}^{(2)}(\bm{k},\bm{k}')$ given in (\ref{Delta_ijkl_1}) and (\ref{Delta_ijkl_2}) are quite general and can be used for inhomogeneous inflation models other than the one discussed in this work.
For later convenience, we also define the power spectra for the polarization modes,
\begin{equation}
\Delta_{ss'}^{(n)}(\bm{k},\bm{k}')\equiv\sum_{i,j}\sum_{k,l}e_{ij}^{\left(s\right)\ast}(\hat{\bm{k}})e_{kl}^{\left(s'\right)\ast}(\hat{\bm{k}}')\Delta_{ij,kl}^{(n)}\left(\bm{k},\bm{k}'\right),\qquad n=1,2.
\end{equation}

Now we are ready to evaluate (\ref{Delta_ijkl_1}) and (\ref{Delta_ijkl_2}) by plugging the expressions for $\mathcal{C}$'s and $\mathcal{D}$'s in our model.
In general, one has to solve the inhomogeneous background $X_{ij}(\bm{x})$ and $\bar{\phi}^{I}(\bm{x})$ from the background equations of motion discussed in Sec.\ref{sec:bg}.
In a general setting, these equations are generally highly non-linear coupled partial differential equations, which are complicated to deal with.
A complete treatment is thus out of the scope of this work.
In this work, we will focus on these corrections under several special configurations of the background inhomogeneities.

\subsection{Contributions from background inhomogeneities} \label{sec:cbi_gw}

Here we evaluate $\Delta_{ij,kl}^{(1)}(\bm{k},\bm{k}')$ given in (\ref{Delta_ijkl_1}).
Instead of making a general ansatz for the inhomogeneous configuration of $X_{ij}(\bm{x})$, we make an irreducible decomposition for $X_{ij}$ similar in (\ref{Hij_dec}), but in the homogeneous background,
\begin{equation}
X_{ij}=2\alpha\,\delta_{ij}+2\left(\partial_{i}\partial_{j}-\frac{1}{3}\delta_{ij}\partial^{2}\right)\beta+\partial_{i}\xi_{j}+\partial_{j}\xi_{i}+\eta_{ij},\label{Xij_dec}
\end{equation}
with $\partial_{i}\xi_{i} = \partial_{i}\eta_{ij} = \eta_{ii} = 0$.
In the following, we evaluate the contributions to  $\Delta_{ij,kl}^{(1)}(\bm{k},\bm{k}')$ and $\Delta_{ij,kl}^{(2)}(\bm{k},\bm{k}')$ from $\alpha$, $\beta$, $\xi_{i}$, $\eta_{ij}$ and $\delta\phi^{I}$ separately.

\subsubsection{Contribution from $\alpha$}

In the simplest case, $X_{ij} \supset 2\alpha \delta_{ij}$, and the corresponding inhomogeneity in the background metric $\bar{h}_{ij}$ is of scalar type.
Since $X_{ij}(\bm{k})$ must satisfy (\ref{Xphi_rc}) in order to ensure that all $\mathcal{C}$'s and $\mathcal{D}$'s factors (\ref{Ham_gg_1})-(\ref{Ham_gf_1}) are real, in Fourier space $\alpha(\bm{k})$ is also subject to the condition $\alpha^{\ast}\left(\bm{k}\right)=\alpha\left(-\bm{k}\right)=\alpha\left(\bm{k}\right)$.
The contributions to $\mathcal{C}_{ij,kl}$ and $\mathcal{D}_{ij,kl}$ from $X_{ij}^{(\alpha)} \equiv 2\alpha \delta_{ij}$ are given in (\ref{calC_ijkl_a}) and (\ref{calD_ijkl_a}), respectively.
Plugging these results into (\ref{Delta_ijkl_1}), the leading order correction to the power spectra of circular polarization modes of the tensor perturbations from $X_{ij}^{(\alpha)}$ are given by:
	\begin{eqnarray}
	\Delta_{ss'}^{(1),\alpha}\left(\bm{k},\bm{k}'\right) & = & -\bigg\{ e_{ij}^{\left(s\right)\ast}(\hat{\bm{k}})e_{ij}^{\left(s'\right)\ast}(\hat{\bm{k}}')\left[1+3(\hat{\bm{k}}\cdot\hat{\bm{k}}')\left(1+\frac{k}{k'}+\frac{k'}{k}\right)\right]\nonumber \\
	&  & -6e_{im}^{\left(s\right)\ast}(\hat{\bm{k}})e_{jm}^{\left(s'\right)\ast}(\hat{\bm{k}}')\hat{\bm{k}}'_{i}\hat{\bm{k}}_{j}\left(1+\frac{k}{k'}+\frac{k'}{k}\right)\bigg\}\frac{2H^{2}}{kk'\left(k+k'\right)}\alpha(\bm{k}+\bm{k}').\label{Delta_ssp_1_a}
	\end{eqnarray}
	
The two different vectors $\bm{k}$ and $\bm{k}'$ in three-dimensional space are coplanar. Without loss of generality  we choose the Cartesian coordinates in the three-dimensional space such that the vectors $\bm{k}$ and $\bm{k}'$ are in the $x$-$z$ plane, 
In particular, we assume $\hat{\bm{k}}=(0,0,1)$ and $\hat{\bm{k}}'=(\sin\theta,0,\cos\theta)$ in Cartesian coordinates, with $0<\theta\leq \pi$.
With the above convention, the polarization tensors can be evaluated explicitly, 
	\begin{equation}
	e_{ij}^{(\pm2)}(\hat{\bm{k}})=\frac{1}{2}\left(\begin{array}{ccc}
		1 & \pm i & 0\\
		\pm i & -1 & 0\\
		0 & 0 & 0
	\end{array}\right),
\end{equation}
and
	\begin{equation}
	e_{ij}^{(\pm2)}(\hat{\bm{k}}')=\frac{1}{2}\left(\begin{array}{ccc}
	\frac{1}{2}(\cos(2\theta)+1) & \pm i\cos\theta & -\frac{1}{2}\sin(2\theta)\\
	\pm i\cos\theta & -1 & \mp i\sin\theta\\
	-\frac{1}{2}\sin(2\theta) & \mp i\sin\theta & \frac{1}{2}(1-\cos(2\theta))
	\end{array}\right). \label{eij_theta}
	\end{equation}
which yield
	\begin{eqnarray}
	e_{ij}^{(s)\ast}(\hat{\bm{k}})e_{ij}^{(s')\ast}(\hat{\bm{k}}') & = & \sin^{4}\left(\frac{\theta}{2}\right),\qquad s=s', \label{eijeij_1} \\
	e_{ij}^{(s)\ast}(\hat{\bm{k}})e_{ij}^{(s')\ast}(\hat{\bm{k}}') & = & \cos^{4}\left(\frac{\theta}{2}\right),\qquad s\neq s',
	\end{eqnarray}
	and
	\begin{eqnarray}
	e_{im}^{(s)\ast}(\hat{\bm{k}})e_{jm}^{(s')\ast}(\hat{\bm{k}}')\hat{\bm{k}}'_{i}\hat{\bm{k}}_{j} & = & \frac{1}{2}\sin^{2}\left(\frac{\theta}{2}\right)\sin^{2}\theta, \qquad s=s'\\
	e_{im}^{(s)\ast}(\hat{\bm{k}})e_{jm}^{(s')\ast}(\hat{\bm{k}}')\hat{\bm{k}}'_{i}\hat{\bm{k}}_{j} & = & -\frac{1}{8}\sin^{4}\theta\csc^{2}\left(\frac{\theta}{2}\right),\qquad s\neq s'. \label{eijeij_4}
	\end{eqnarray}

Using these results, the power spectra for the polarization modes (\ref{Delta_ssp_1_a}) can be written as
	\begin{eqnarray}
	&  & \Delta_{++}^{(1),\alpha}\left(\bm{k},\bm{k}'\right)=\Delta_{--}^{(1),\alpha}\left(\bm{k},\bm{k}'\right)\nonumber \\
	& = & \frac{H^{2}}{k^{3}}\sin^{4}\left(\frac{\theta}{2}\right)\left[6+5r+6r^{2}+3\left(1+r+r^{2}\right)\cos\theta\right]\frac{2}{r^{2}\left(1+r\right)}\alpha\left(\bm{k}+\bm{k}'\right),\label{Delta_1_pp_a}
	\end{eqnarray}
	and
	\begin{eqnarray}
	&  & \Delta_{+-}^{(1),\alpha}\left(\bm{k},\bm{k}'\right)=\Delta_{-+}^{(1),\alpha}\left(\bm{k},\bm{k}'\right)\nonumber \\
	& = & \frac{H^{2}}{k^{3}}\cos^{4}\left(\frac{\theta}{2}\right)\left[-6-7r-6r^{2}+3\left(1+r+r^{2}\right)\cos\theta\right]\frac{2}{r^{2}\left(1+r\right)}\alpha\left(\bm{k}+\bm{k}'\right),\label{Delta_1_pm_a}
	\end{eqnarray}
where $\cos\theta\equiv\hat{\bm{k}}\cdot\hat{\bm{k}}'$ and we defined 
	\begin{equation}
	 r\equiv\frac{k'}{k},\label{theta_r_def}
	\end{equation}
for short.
We can draw several conclusions from (\ref{Delta_1_pp_a}) and (\ref{Delta_1_pm_a}).
\begin{itemize}
	\item In our model both isotropy and homogeneity are broken, and thus the power spectra do not take the standard form. In particular, there is no $\delta^3(\bm{k}+\bm{k}')$ factor in the above expressions.
	\item In the case of $X_{ij} \supset 2\alpha \delta_{ij}$, the two circular polarization modes acquire the same amount of powers as $\Delta_{++}^{(1),\alpha}=\Delta_{--}^{(1),\alpha}$.
	\item (\ref{Delta_1_pm_a}) implies that, since the spatial homogeneity is broken, there are correlations between different Fourier modes, and thus there is non-vanishing correlation between the two circular polarization modes as $\Delta_{+-}^{(1),\alpha} = \Delta_{-+}^{(1),\alpha} \neq 0$.
\end{itemize}

As a consistency check, let us consider the homogeneous limit when $\alpha(\bm{x}) \rightarrow \alpha_{0}$, which is a small constant. In this limit,
	\begin{equation}
		\alpha(\bm{k}+ \bm{k}') \rightarrow \alpha_{0} (2\pi)^3 \delta^3(\bm{k}+\bm{k}'),
	\end{equation}
which also implies $\bm{k}'\rightarrow -\bm{k}$ and thus $r\rightarrow 1$, $\theta\rightarrow \pi$.
In this homogeneous limit, 
	\begin{equation}
	\Delta_{++}^{(1),\alpha}=\Delta_{--}^{(1),\alpha}\rightarrow-8\frac{H^{2}}{k^{3}}\alpha_{0}\left(2\pi\right)^{3}\delta^{3}\left(\bm{k}+\bm{k}'\right),
	\end{equation}
which takes the standard form and corresponds to nothing but rescaling of the scale factor. (\ref{Delta_1_pm_a}) becomes
	\begin{equation}
	\Delta_{+-}^{(1),\alpha}=\Delta_{-+}^{(1),\alpha}\rightarrow 0,
	\end{equation}
which implies the two circular polarization modes are uncorrelated in the homogeneous limit, as expected.

\subsubsection{Contribution from $\beta$}

In this case, $X_{ij}^{(\beta)}\left(\bm{k}\right)=2\left(-\bm{k}_{i}\bm{k}_{j}+\frac{1}{3}\delta_{ij}k^{2}\right)\beta\left(\bm{k}\right)$ and the calculation is parallel to the above. 
Note $\beta(\bm{k})$ must satisfy the same condition as that of $\alpha(\bm{k})$, i.e., $\beta^{\ast}\left(\bm{k}\right)=\beta\left(-\bm{k}\right)=\beta\left(\bm{k}\right)$ in order to ensure $\mathcal{C}$'s and $\mathcal{D}$'s are real. The factors in (\ref{calC_ijkl}) and (\ref{calD_ijkl}) become (\ref{calC_ijkl_b}) and (\ref{calD_ijkl_b}). 
The leading order corrections to the power spectra of tensor modes from $X_{ij}^{(\beta)}$ are:
	\begin{eqnarray}
	&  & \Delta_{ss'}^{(1),\beta}\left(\bm{k},\bm{k}'\right)\nonumber \\
	& = & \frac{H^{2}}{k+k'}\bigg\{4e_{ij}^{(s)\ast}(\hat{\bm{k}})e_{ij}^{(s')\ast}(\hat{\bm{k}}')\left[\frac{2}{3}-\frac{4}{3}\hat{\bm{k}}\cdot\hat{\bm{k}}'+\left(\frac{1}{3}-(\hat{\bm{k}}\cdot\hat{\bm{k}}')^{2}\right)\left(1+\frac{k}{k'}+\frac{k'}{k}\right)\right]\nonumber \\
	&  & -8e_{im}^{(s)\ast}(\hat{\bm{k}})e_{jm}^{(s')\ast}(\hat{\bm{k}}')\hat{\bm{k}}'_{i}\hat{\bm{k}}_{j}\left[\frac{k^{2}}{k'^{2}}+\frac{k'^{2}}{k^{2}}+(1+\hat{\bm{k}}\cdot\hat{\bm{k}}')\left(1+\frac{k}{k'}+\frac{k'}{k}\right)\right]\bigg\}\beta(\bm{k}+\bm{k}'). \label{Delta_ssp_1_b}
	\end{eqnarray}
Using the representation of the polarization tensors (\ref{eijeij_1})-(\ref{eijeij_4}), we get
	\begin{eqnarray}
	&  & \Delta_{++}^{(1),\beta}\left(\bm{k},\bm{k}'\right)=\Delta_{--}^{(1),\beta}\left(\bm{k},\bm{k}'\right)\nonumber \\
	& = & -\frac{H^{2}}{k}\sin^{4}\left(\frac{\theta}{2}\right)\big[12+19r+15r^{2}+19r^{3}+12r^{4}+9r\left(1+r+r^{2}\right)\cos\left(2\theta\right)\nonumber \\
	&  & +4\left(3+6r+8r^{2}+6r^{3}+3r^{4}\right)\cos\theta\big]\frac{2}{3r^{2}\left(1+r\right)}\beta\left(\bm{k}+\bm{k}'\right),\label{Delta_1_pp_b}
	\end{eqnarray}
	and
	\begin{eqnarray}
	&  & \Delta_{+-}^{(1),\beta}\left(\bm{k},\bm{k}'\right)=\Delta_{-+}^{(1),\beta}\left(\bm{k},\bm{k}'\right)\nonumber \\
	& = & \frac{H^{2}}{k}\cos^{4}\left(\frac{\theta}{2}\right)\big[12+5r+9r^{2}+5r^{3}+12r^{4}-4\left(3+2r^{2}+3r^{4}\right)\cos\theta\nonumber \\
	&  & -9r\left(1+r+r^{2}\right)\cos\left(2\theta\right)\big]\frac{2}{3r^{2}\left(1+r\right)}\beta\left(\bm{k}+\bm{k}'\right).\label{Delta_1_pm_b}
	\end{eqnarray}
Again, the two circular polarization modes have the same power and non-vanishing cross-correlations.

In the homogeneous limit with $\beta(\bm{x})\rightarrow \beta_{0} = \mathrm{const}$ (and thus $r\rightarrow1$, $\theta\rightarrow \pi$), we have
	\begin{equation}
	\Delta_{ss'}^{(1),\beta} \rightarrow 0, \qquad s,s'=\pm 2,
	\end{equation}
which are consistent with the fact in this limit the background metric becomes exactly $\bar{h}_{ij} \rightarrow \delta_{ij}$ and there is no correction to the power spectra at all.

\subsubsection{Contribution from $\xi_{i}$} \label{sec:cf_xi_gw}

In this case, the inhomogeneity in the background metric takes the vector form. 
In Fourier space $X_{ij}^{(\xi)}\left(\bm{k}\right)=i\bm{k}_{i}\xi_{j}\left(\bm{k}\right)+i\bm{k}_{j}\xi_{i}\left(\bm{k}\right)$, where $\bm{\xi}_{i}\left(\bm{k}\right)$ must be purely imaginary and satisfies
	\begin{equation}
	\xi_{i}^{\ast}\left(\bm{k}\right)=\xi_{i}\left(-\bm{k}\right)=-\xi_{i}\left(\bm{k}\right),\label{xi_cond}
	\end{equation}
in order to ensure that $X_{ij}^{(\xi)}(\bm{k})$ is real.
Plugging the expression for $X_{ij}^{(\xi)}$
into (\ref{calC_ijkl}) and (\ref{calD_ijkl}), we get(\ref{calC_ijkl_xi}) and (\ref{calD_ijkl_xi}). 
The leading order correction to the power spectra of tensor modes from $X_{ij}^{(\xi)}$ thus reads
	\begin{eqnarray}
	\Delta_{ss'}^{(1),\xi}\left(\bm{k},\bm{k}'\right) & = & 2\frac{H^{2}}{k'k(k'+k)}\bigg[\frac{k^{4}+k^{3}k'+kk'^{3}+k'^{4}}{k^{2}k'^{2}}\nonumber \\
	&  & \qquad\times\left(e_{im}^{\left(s\right)\ast}(\hat{\bm{k}})e_{jm}^{\left(s'\right)\ast}(\hat{\bm{k}}')\bm{k}_{j}+e_{jm}^{\left(s\right)\ast}(\hat{\bm{k}})e_{im}^{\left(s'\right)\ast}(\hat{\bm{k}}')\bm{k}'_{j}\right)i\xi_{i}\left(\bm{k}+\bm{k}'\right)\nonumber \\
	&  & \quad-e_{ij}^{\left(s\right)\ast}(\hat{\bm{k}})e_{ij}^{\left(s'\right)\ast}(\hat{\bm{k}}')i\bm{k}\cdot\bm{\xi}\left(\bm{k}+\bm{k}'\right)\left(k'^{2}-k^{2}\right)\frac{k'^{2}+k'k+k^{2}}{k'^{2}k^{2}}\bigg].\label{Delta_ssp_1_xi}
	\end{eqnarray}
	
In order to evaluate the power spectra (\ref{Delta_ssp_1_xi}) explicitly, we decompose $\bm{\xi}_{i}$ into two polarization modes.
For $\bm{k}$ along $\left(\theta,\phi\right)$-direction,
the circular polarization vectors with $\lambda=\pm 1$ are given by
\begin{equation}
\bm{e}_{i}^{(\lambda)}(\hat{\bm{k}})=\frac{1}{\sqrt{2}}\left(\begin{array}{c}
\cos\theta\cos\phi-\lambda i\sin\phi\\
\cos\theta\sin\phi+\lambda i\cos\phi\\
-\sin\theta
\end{array}\right),\label{epem}
\end{equation}
which satisfy
\begin{equation}
e_{i}^{(\lambda)\ast}(\bm{k})=e_{i}^{(\lambda)}(-\bm{k})=e_{i}^{(-\lambda)}(\bm{k}).\label{ei_cond}
\end{equation}
Since $\bm{\xi}_{i}$ itself is purely imaginary, it is convenient
to write
\begin{equation}
\xi_{i}(\bm{k})=i\tilde{\xi}_{i}(\bm{k})=\sum_{\lambda=\pm1}i\tilde{\xi}^{(\lambda)}(\bm{k})e_{i}^{(\lambda)}(\bm{k}),\label{xi_dec}
\end{equation}
where the polarization modes $\tilde{\xi}^{(\lambda)}(\bm{k})$ must satisfy 
\begin{equation}
\tilde{\xi}^{(\lambda)\ast}(\bm{k})=-\tilde{\xi}^{(\lambda)}(-\bm{k})=\tilde{\xi}^{(-\lambda)}(\bm{k}),\label{xi_lambda_cond}
\end{equation}
as a result of (\ref{xi_cond}).
	
With our convention, $\bm{k}+\bm{k}'$ is given by
\begin{equation}
\bm{k}+\bm{k}'=\left|\bm{k}+\bm{k}'\right|\left(\sin\rho,0,\cos\rho\right),\label{kpluskp}
\end{equation}
with
\begin{equation}
\left|\bm{k}+\bm{k}'\right|=\sqrt{k^{2}+k'^{2}+2kk'\cos\theta},
\end{equation}
and
\begin{eqnarray}
\sin\rho & = & \frac{k'\sin\theta}{\sqrt{k^{2}+k'^{2}+2kk'\cos\theta}},\label{sin_varphi}\\
\cos\rho & = & \frac{k'\cos\theta+k}{\sqrt{k^{2}+k'^{2}+2kk'\cos\theta}}.\label{cos_varphi}
\end{eqnarray}
The circular polarization vectors are thus
\begin{equation}
\bm{e}_{i}^{(\pm)}\left(\bm{k}+\bm{k}'\right)=\frac{1}{\sqrt{2}}\left(\begin{array}{c}
\cos\rho\\
\pm i\\
-\sin\rho
\end{array}\right).
\end{equation}

Using these results, the corrections to the power spectra for the polarization modes of tensor perturbations are 
	\begin{eqnarray}
	\Delta_{++}^{(1),\xi}\left(\bm{k},\bm{k}'\right) & = & \frac{H^{2}(r-1)}{\sqrt{2}k^{2}r^{3}}\sin^{2}\left(\frac{\theta}{2}\right)\sin\theta\Big[\left(r^{3}+1\right)x_{-}(\bm{k}+\bm{k}')\nonumber \\
	&  & \quad+\frac{r^{4}-r^{2}+1+\left(r^{2}+r+1\right)r\cos\theta}{\sqrt{r^{2}+2r\cos\theta+1}}x_{+}(\bm{k}+\bm{k}')\Big],
	\end{eqnarray}
with
	\begin{equation}
	x_{\pm}\equiv\tilde{\xi}^{(+)}\pm\tilde{\xi}^{(-)},\label{xpm_def}
	\end{equation}
and
	\begin{eqnarray}
	&  & \Delta_{++}^{(1),\xi}\left(\bm{k},\bm{k}'\right)-\Delta_{--}^{(1),\xi}\left(\bm{k},\bm{k}'\right)\nonumber \\
	& = & \frac{\sqrt{2}H^{2}}{k^{2}r^{3}}\left(r^{4}-r^{3}+r-1\right)\sin^{2}\left(\frac{\theta}{2}\right)\sin\theta\,x_{-}(\bm{k}+\bm{k}'). \label{Delta_1_xi_ppmm}
	\end{eqnarray}
(\ref{Delta_1_xi_ppmm}) implies that the two circular polarization modes for tensor perturbations have different power, if the two circular polarization modes for the background vector $\xi_{i}(\bm{k})$ have different amplitudes, i.e., $\tilde{\xi}^{(+)}\neq\tilde{\xi}^{(-)}$ and thus $x_{-}\neq 0$. 
We also have
	\begin{eqnarray}
	\Delta_{+-}^{(1),\xi}\left(\bm{k},\bm{k}'\right) & = & \frac{H^{2}}{\sqrt{2}k^{2}r^{3}}\cos^{2}\left(\frac{\theta}{2}\right)\sin\theta\Big[(r+1)^{2}\left(r^{2}-r+1\right)x_{-}\left(\bm{k}+\bm{k}'\right)\nonumber \\
	&  & \quad-\frac{r^{5}+r^{4}-r^{3}+r^{2}-r-1+\left(r^{3}-1\right)r\cos\theta}{\sqrt{r^{2}+2r\cos\theta+1}}x_{+}(\bm{k}+\bm{k}')\Big],
	\end{eqnarray}
and
	\begin{eqnarray}
	&  & \Delta_{+-}^{(1),\xi}\left(\bm{k},\bm{k}'\right)-\Delta_{-+}^{(1),\xi}\left(\bm{k},\bm{k}'\right)\nonumber \\
	& = & \frac{\sqrt{2}H^{2}}{k^{2}r^{3}}\left(r^{4}+r^{3}+r+1\right)\cos^{2}\left(\frac{\theta}{2}\right)\sin\theta\,x_{-}(\bm{k}+\bm{k}'). \label{Delta_1_xi_pmmp}
	\end{eqnarray}
Again, $\Delta_{+-}^{(1),\xi}\neq\Delta_{-+}^{(1),\xi}$ when $x_{-}\neq 0$.

As a consistency check, let us consider the above expressions in the homogeneous limit, which corresponds to $\xi_{i}(\bm{x}) \rightarrow  \xi_{0i}$ with $\xi_{0i}$ a constant (spatially homogeneous) vector field.
In this limit, $x_\pm$ are constant numbers, and thus $x_{\pm}(\bm{k}+\bm{k}') \propto \delta^3(\bm{k}+\bm{k}')$, which implies $r\rightarrow 0$ and $\theta \rightarrow  \pi$.
As a result,
	\begin{equation}
		\Delta_{ss'}^{(1),\xi} \rightarrow 0,\qquad s,s'=\pm 2,
	\end{equation}
and thus there is no correction to the power spectra, which is consistent with the fact that the background spatial metric becomes exactly $\delta_{ij}$ in this limit.

\subsubsection{Contribution from $\eta_{ij}$} \label{sec:cf_eta_gw}

In this case, the inhomogeneity in the background metric takes the tensor form, which can be viewed as a classical gravitational waves background. In Fourier space $X_{ij}^{(\eta)}(\bm{k})=\eta_{ij}(\bm{k})$, where $\eta_{ij}(\bm{k})$ must satisfy $\eta_{ij}^{\ast}(\bm{k})=\eta_{ij}(-\bm{k})=\eta_{ij}(\bm{k})$.
and we have (\ref{calC_ijkl_eta}) and (\ref{calD_ijkl_eta}). 
The leading order corrections to the power spectra of circular polarization modes of the tensor perturbations from $X_{ij}^{(\eta)}$ are given by:
	\begin{eqnarray}
	\Delta_{ss'}^{(1),\eta}\left(\bm{k},\bm{k}'\right) & = & -\frac{2H^{2}}{k'k(k'+k)}\bigg\{2e_{im}^{\left(s\right)\ast}(\hat{\bm{k}})e_{jm}^{\left(s'\right)\ast}(\hat{\bm{k}}')\eta_{ij}\left(\bm{k}+\bm{k}'\right)\nonumber \\
	&  & \qquad\times\left[1+\left(1+\frac{k'}{k}+\frac{k}{k'}\right)\left(\frac{k}{k'}+\frac{k'}{k}+3\hat{\bm{k}}\cdot\hat{\bm{k}}'\right)\right]\nonumber \\
	&  & \quad+\left(1+\frac{k'}{k}+\frac{k}{k'}\right)e_{ij}^{\left(s\right)\ast}(\hat{\bm{k}})e_{ij}^{\left(s'\right)\ast}(\hat{\bm{k}}')\hat{\bm{k}}_{m}\hat{\bm{k}}'_{n}\eta_{mn}\left(\bm{k}+\bm{k}'\right)\bigg\}.\label{Delta_ssp_1_eta}
	\end{eqnarray}

In order to evaluate (\ref{Delta_ssp_1_eta}) explicitly, we make the same decomposition for the background tensorial inhomogeneity $\eta_{ij}$
	\begin{equation}
	\eta_{ij}(\bm{k})=\sum_{s=\pm2}\eta^{(s)}(\bm{k})e_{ij}^{(s)}(\bm{k}), \label{eta_dec}
	\end{equation}
to that of the tensor perturbations $\gamma_{ij}$, where the polarization modes $\eta^{(s)}(\bm{k})$ satisfy $\eta^{(s)\ast}(\bm{k})=\eta^{(s)}(-\bm{k})=\eta^{(-s)}(\bm{k})$. 
With our convention, $\bm{k}+\bm{k}'$ is given in (\ref{kpluskp}), and thus $e_{ij}^{(s)}(\bm{k}+\bm{k}')$ takes the same form as (\ref{eij_theta}) but with $\theta$ replaced by $\rho$.
With these results, the corrections to the power spectra for the tensor polarization modes  are 
	\begin{equation}
	\Delta_{++}^{(1),\eta}\left(\bm{k},\bm{k}'\right)=\frac{H^{2}}{k^{3}}\frac{\sin^{2}\theta}{2r^{3}\sqrt{r^{2}+2r\cos\theta+1}}\left(\mathcal{A}_{+}y_{+}(\bm{k}+\bm{k}')+\mathcal{A}_{-}y_{-}(\bm{k}+\bm{k}')\right),\label{Delta_1_eta_pp}
	\end{equation}
	with 
	\begin{equation}
	y_{\pm}\equiv\eta^{(+2)}\pm\eta^{(-2)}, \label{ypm_def}
	\end{equation}
	and
\begin{eqnarray}
\mathcal{A}_{+} & \equiv & \frac{1}{4(r+1)\sqrt{r^{2}+2r\cos\theta+1}}\big[4r^{6}+8r^{5}+29r^{4}+29r^{3}+29r^{2}+8r+4\nonumber \\
&  & \quad+7\left(r^{2}+r+1\right)r^{2}\cos(2\theta)+4\left(4r^{4}+6r^{3}+11r^{2}+6r+4\right)r\cos\theta\big],\label{calA_p}
\end{eqnarray}
and
\begin{eqnarray}
\mathcal{A}_{-} & \equiv & r^{4}+r^{3}+3r^{2}+r+1+3\left(r^{2}+r+1\right)r\cos\theta\label{calA_m}
\end{eqnarray}
Generally, $\Delta_{--}^{(1),\eta}\neq\Delta_{++}^{(1),\eta}$ and satisfy  
	\begin{equation}
	\Delta_{++}^{(1),\eta}\left(\bm{k},\bm{k}'\right)-\Delta_{--}^{(1),\eta}\left(\bm{k},\bm{k}'\right)=\frac{H^{2}\sin^{2}\theta}{k^{3}r^{3}\sqrt{r^{2}+2r\cos\theta+1}}\mathcal{A}_{-}y_{-}(\bm{k}+\bm{k}'),
	\end{equation}
with $\mathcal{A}_{-}$ given in (\ref{calA_m}).
Similar to the case of vector-form background inhomogeneity $\xi_{i}$, the two circular polarization modes of gravitational waves $\gamma^{(\pm 2)}$ will acquire different power, if the two polarization modes of background tensorial inhomogeneity have different amplitudes, i.e., $\eta^{(+2)}\neq\eta^{(-2)}$ and thus $y_{-}\neq 0$.

The cross-correlation between the two circular polarization modes are not vanishing generally, which is given by
	\begin{equation}
	\Delta_{+-}^{(1),\eta}\left(\bm{k},\bm{k}'\right)=\frac{H^{2}}{k^{3}}\frac{\sin^{2}\theta}{2r^{3}(r+1)\sqrt{r^{2}+2r\cos\theta+1}}\left(\mathcal{B}_{+}y_{+}(\bm{k}+\bm{k}')+\mathcal{B}_{-}y_{-}(\bm{k}+\bm{k}')\right),\label{Delta_1_eta_pm}
	\end{equation}
	with
	\begin{eqnarray}
	\mathcal{B}_{+} & = & \frac{1}{4\sqrt{r^{2}+2r\cos\theta+1}}\big[4r^{6}+21r^{4}+5r^{3}+21r^{2}+4\nonumber \\
	&  & \quad+7\left(r^{2}+r+1\right)r^{2}\cos(2\theta)+4\left(4r^{4}+2r^{3}+7r^{2}+2r+4\right)r\cos\theta\big],\label{calB_p}
	\end{eqnarray}
	and
	\begin{eqnarray}
	\mathcal{B}_{-} & = & -r^{5}-2r^{3}+2r^{2}+1-3\left(r^{3}-1\right)r\cos\theta.\label{calB_m}
	\end{eqnarray}
We also have
	\begin{equation}
	\Delta_{+-}^{(1),\eta}\left(\bm{k},\bm{k}'\right)-\Delta_{-+}^{(1),\eta}\left(\bm{k},\bm{k}'\right) = \frac{H^{2}\sin^{2}\theta}{k^{3}r^{3}(r+1)\sqrt{r^{2}+2r\cos\theta+1}}\mathcal{B}_{-}y_{-}(\bm{k}+\bm{k}'),
	\end{equation}
which is non-vanishing when $\eta^{(+2)}\neq\eta^{(-2)}$.

As a consistency check, let us consider the above expressions in the homogeneous limit, which corresponds to $\eta_{ij}(\bm{x}) \rightarrow  \eta_{0ij}$ with $\eta_{0ij}$ a constant (spatially homogeneous) transverse-traceless field.
In this limit, $y_\pm$ are constant numbers, and thus $y_{\pm}(\bm{k}+\bm{k}') \propto \delta^3(\bm{k}+\bm{k}')$, which implies $r\rightarrow 0$ and $\theta \rightarrow  \pi$.
As a result,
\begin{equation}
\Delta_{ss'}^{(1),\eta} \rightarrow 0,\qquad s,s'=\pm 2,
\end{equation}
and thus there is no correction to the power spectra, which is consistent with the fact that the background spatial metric becomes $\bar{h}_{ij} \supset \delta_{ij}+\eta_{0ij}$ in this limit, which is equivalent to a homogeneous and isotropic one by a global coordinate transformation.

\subsection{Contributions from interactions with $\delta\phi^{I}$} \label{sec:cf_df_gw}

Now let us consider the corrections to the power spectra of the gravitational waves from interactions between tensor perturbations $\gamma_{ij}$ and scalar field perturbations $\delta\phi^{I}$. 
The general form is given in (\ref{Delta_ijkl_2}), from which 
the corrections to the power spectra for the circular polarization modes are
\begin{eqnarray}
\Delta_{ss'}^{(2)}(\bm{k},\bm{k}') & = & e_{ij}^{(s)\ast}(\hat{\bm{k}})e_{kl}^{(s')\ast}(\hat{\bm{k}}')8H^{2}\nonumber \\
&  & \times\int\frac{\mathrm{d}^{3}p}{\left(2\pi\right)^{3}}\mathcal{F}\left(k,k',p\right)\mathcal{D}_{ij}^{I}\left(-\bm{k},-\bm{p},\bm{k}+\bm{p}\right)\mathcal{D}_{kl}^{I}\left(-\bm{k}',\bm{p},\bm{k}'-\bm{p}\right),
\end{eqnarray}
where $\mathcal{F}$ is defined in (\ref{calF}).
In our case, $\mathcal{D}_{ij}^{I}$ is given in (\ref{calDIij_fin}), which implies
\begin{equation}
\mathcal{D}_{ij}^{I}\left(-\bm{k},-\bm{p},\bm{k}+\bm{p}\right)=-\left(k_{i}+p_{i}\right)\left(k_{j}+p_{j}\right)\bar{\phi}^{I}\left(\bm{k}+\bm{p}\right),
\end{equation}
and thus
\begin{eqnarray}
\Delta_{ss'}^{(2)}(\bm{k},\bm{k}') & = & 8H^{2} e_{ij}^{(s)\ast}(\hat{\bm{k}})e_{kl}^{(s')\ast}(\hat{\bm{k}}')\nonumber \\
&  & \times\int\frac{\mathrm{d}^{3}p}{\left(2\pi\right)^{3}}p_{i}p_{j}p_{k}p_{l}\mathcal{F}\left(k,k',p\right)\bar{\phi}^{I}\left(\bm{k}+\bm{p}\right)\bar{\phi}^{I}\left(\bm{k}'-\bm{p}\right).\label{Delta_2_ssp}
\end{eqnarray}
We stress that in Fourier space $\phi^I(\bm{k})$ must satisfy $\phi^{\ast}(\bm{k}) = \phi^{I}(-\bm{k}) = \phi^{I}(\bm{k})$ in order ensure the factor $\mathcal{D}_{ij}^{I}$ to be real.
Generally, since the background values of the scalar fields $\phi^{I}$ are not homogeneous, $\bar{\phi}^{I}(\bm{k})$ is not proportional to $\delta^3(\bm{k})$, the integral in (\ref{Delta_2_ssp}) will not contribute $\delta^3(\bm{k}+\bm{k}')$ any more.

In order to get a glimpse of the effects of $\bar{\phi}^{I}$, we make an ansatz for $\bar{\phi}^{I}(\bm{k})$ such that
	\begin{equation}
	\bar{\phi}^{I}\left(\bm{k}\right)=\frac{1}{2}\phi_{\ast}^{I}\left(2\pi\right)^{3}\left[\delta^{3}\left(\bm{k}+\bm{k}_{\ast}\right)+\delta^{3}\left(\bm{k}-\bm{k}_{\ast}\right)\right],\label{phibar_ansatz}
	\end{equation}
with constant $\phi_{\ast}^{I}$ and $k_{\ast}$, which corresponds to the configuration $\bar{\phi}^{I}\left(\bm{x}\right)=\phi_{\ast}^{I}\cos\left(\bm{k}_{\ast}\cdot\bm{x}\right)$. 
In this case, the integral in (\ref{Delta_2_ssp}) can be easily evaluated.
After some manipulations, we find
	\begin{eqnarray}
	\Delta_{ss'}^{(2)}\left(\bm{k},\bm{k}'\right) & = & \frac{H^{2}}{k^{3}}\phi_{\ast}^{I}\phi_{\ast}^{I}\left(2\pi\right)^{3}\big[\delta^{3}\left(\bm{k}+\bm{k}'+2\bm{k}_{\ast}\right)\tilde{\mathcal{F}}_{2}\left(\bm{k},\bm{k}_{\ast}\right)\nonumber \\
	&  & \quad+\delta^{3}\left(\bm{k}+\bm{k}'\right)\tilde{\mathcal{F}}_{0}\left(\bm{k},\bm{k}_{\ast}\right)+\delta^{3}\left(\bm{k}+\bm{k}'-2\bm{k}_{\ast}\right)\tilde{\mathcal{F}}_{-2}\left(\bm{k},\bm{k}_{\ast}\right)\big], \label{Delta_2_ssp_sc}
	\end{eqnarray}
with
	\begin{eqnarray}
	\tilde{\mathcal{F}}_{2} & = & 2k^{3}e_{ij}^{(s)\ast}(\bm{k})e_{kl}^{(s')}(\bm{k}+2\bm{k}_{\ast})\bm{k}_{\ast i}\bm{k}_{\ast j}\bm{k}_{\ast k}\bm{k}_{\ast l}\mathcal{F}\left(k,\left|\bm{k}+2\bm{k}_{\ast}\right|,\left|\bm{k}+\bm{k}_{\ast}\right|\right), \label{tildecalF_p2}\\
	\tilde{\mathcal{F}}_{0} & = & 2k^{3}e_{ij}^{(s)\ast}(\bm{k})e_{kl}^{(s')}(\bm{k})\bm{k}_{\ast i}\bm{k}_{\ast j}\bm{k}_{\ast k}\bm{k}_{\ast l}\left(\mathcal{F}\left(k,k,\left|\bm{k}+\bm{k}_{\ast}\right|\right)+\mathcal{F}\left(k,k,\left|\bm{k}-\bm{k}_{\ast}\right|\right)\right), \label{tildecalF_0}\\
	\tilde{\mathcal{F}}_{-2} & = & 2k^{3}e_{ij}^{(s)\ast}(\bm{k})e_{kl}^{(s')}(\bm{k}-2\bm{k}_{\ast})\bm{k}_{\ast i}\bm{k}_{\ast j}\bm{k}_{\ast k}\bm{k}_{\ast l}\mathcal{F}\left(k,\left|\bm{k}-2\bm{k}_{\ast}\right|,\left|\bm{k}-\bm{k}_{\ast}\right|\right), \label{tildecalF_m2}
	\end{eqnarray}
which are functions of $\bm{k}$ and $\bm{k}_{\ast}$.
With this simple configuration, power spectra for the gravitational waves receive corrections when $\bm{k}+\bm{k}' = 0,\pm 2\bm{k}_{\ast}$.
It is interesting to note among the three contributions in (\ref{Delta_2_ssp_sc}), there is one proportional to $\delta^{3}\left(\bm{k}+\bm{k}'\right)$, which takes the same form of spectrum in anisotropic inflation in which the spatial isotropy is lost while the  homogeneity is preserved.

\begin{figure}[h]
	\begin{minipage}{\textwidth}
		\centering
			\includegraphics[width=0.48\textwidth]{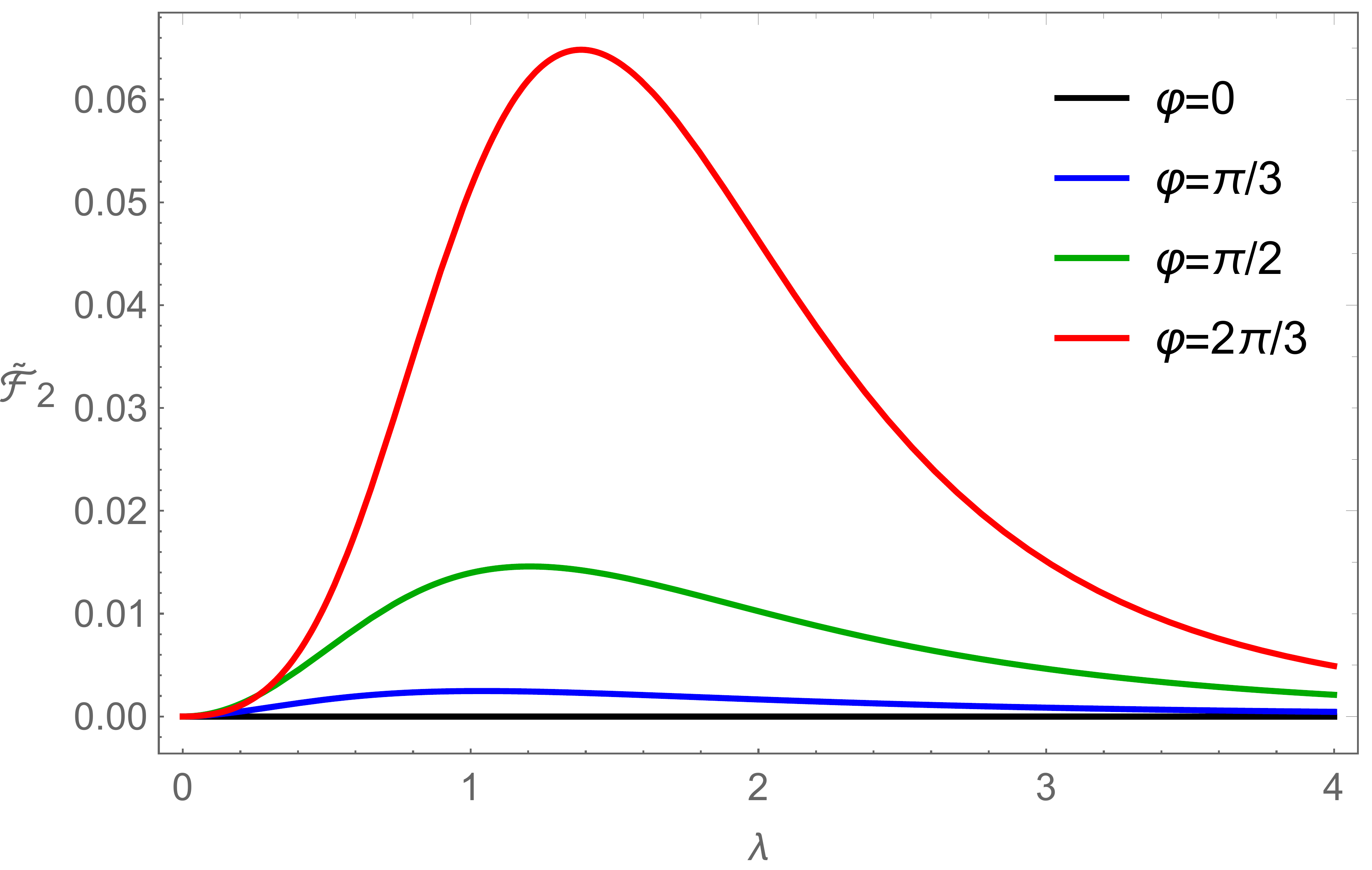}
			\includegraphics[width=0.48\textwidth]{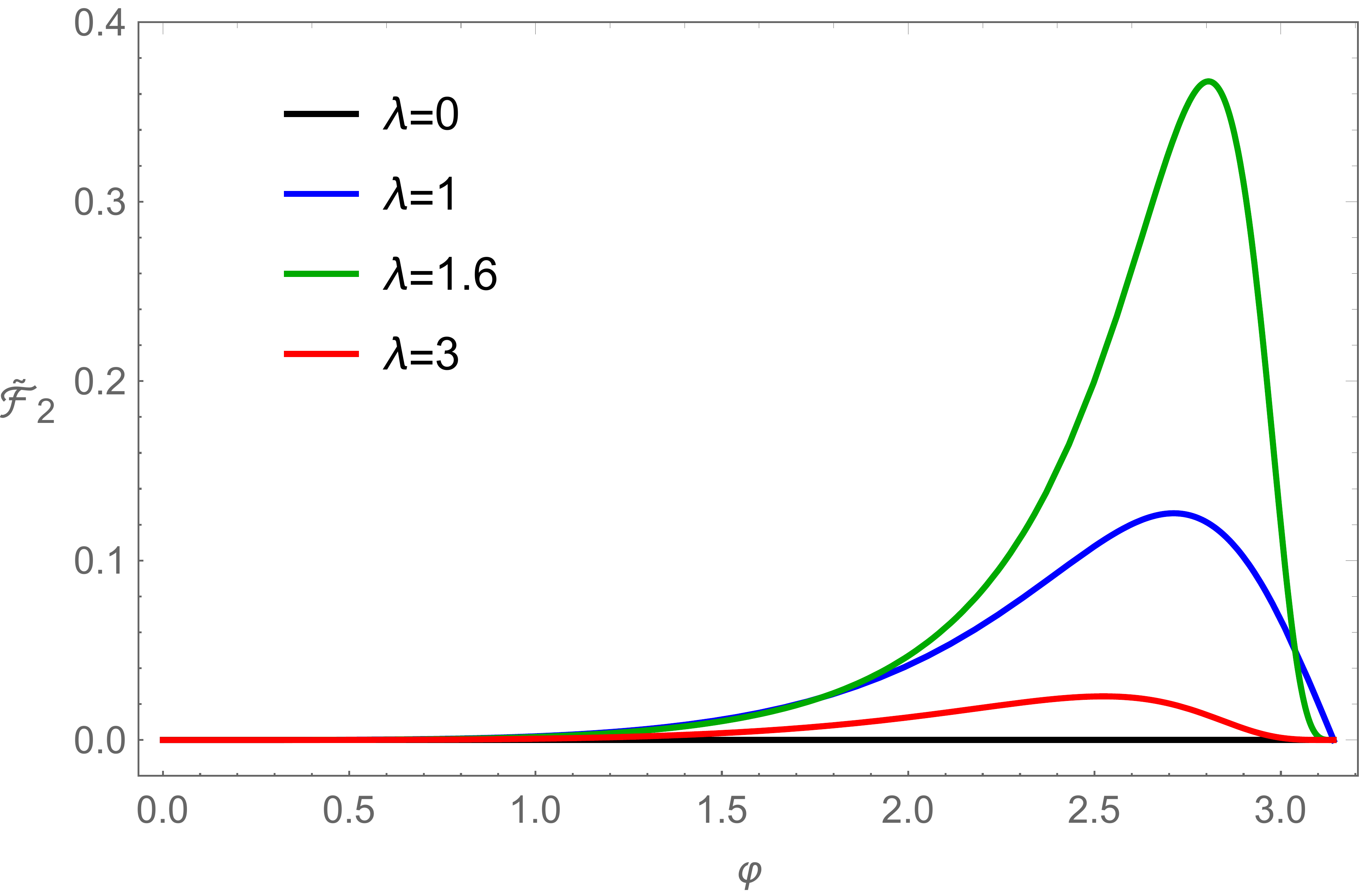}
			\caption{The dependence of $\tilde{F}_{2}(\lambda,\varphi)$ of $\lambda$ with fixed values of $\varphi$ (left pane), and of $\varphi$ with fixed values of $\lambda$ (right pane).}
			\label{fig:Fp2}
	\end{minipage}
\end{figure}

\begin{figure}[h]
	\begin{minipage}{\textwidth}
		\centering
		\includegraphics[width=0.48\textwidth]{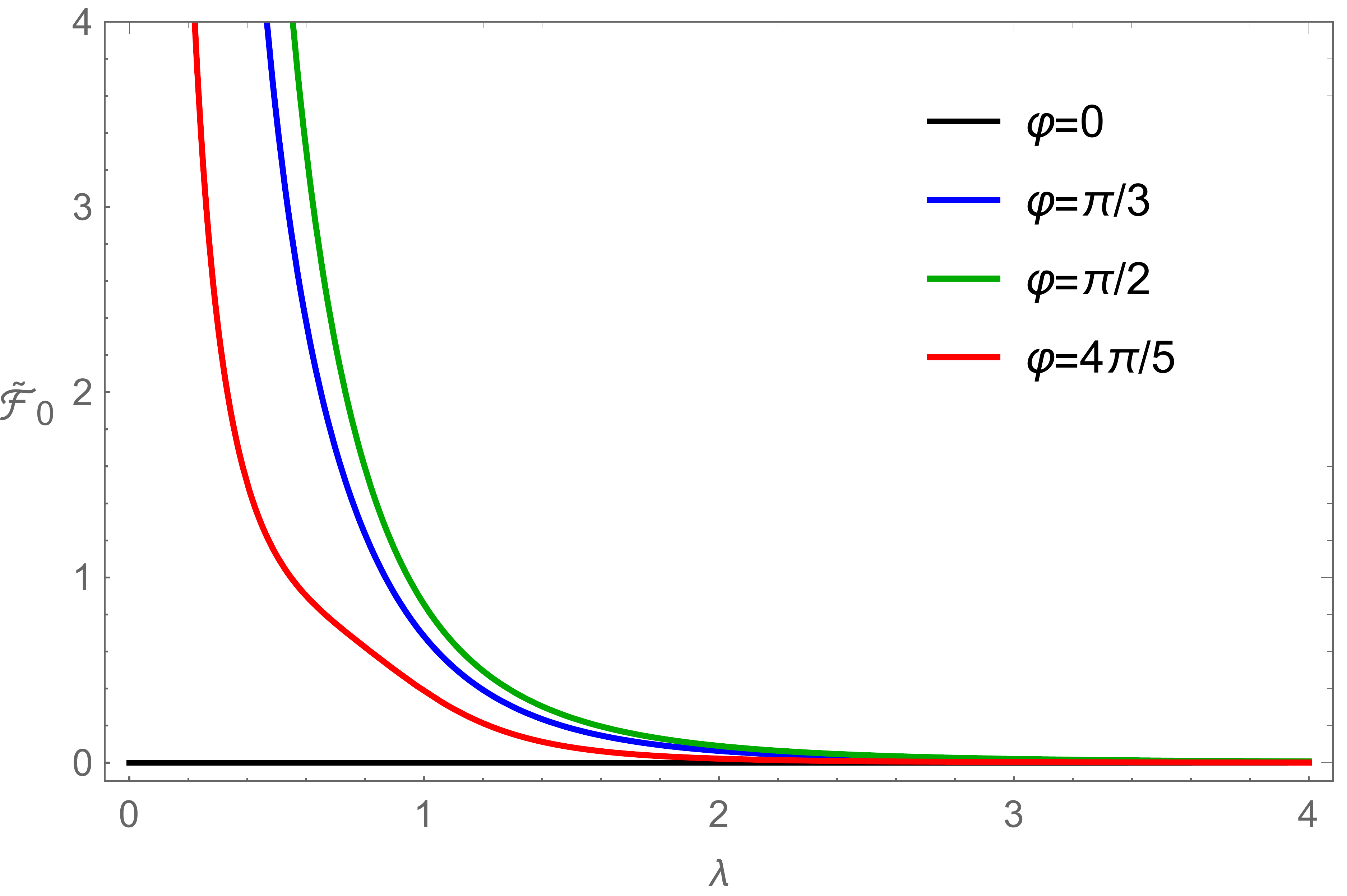}
		\includegraphics[width=0.48\textwidth]{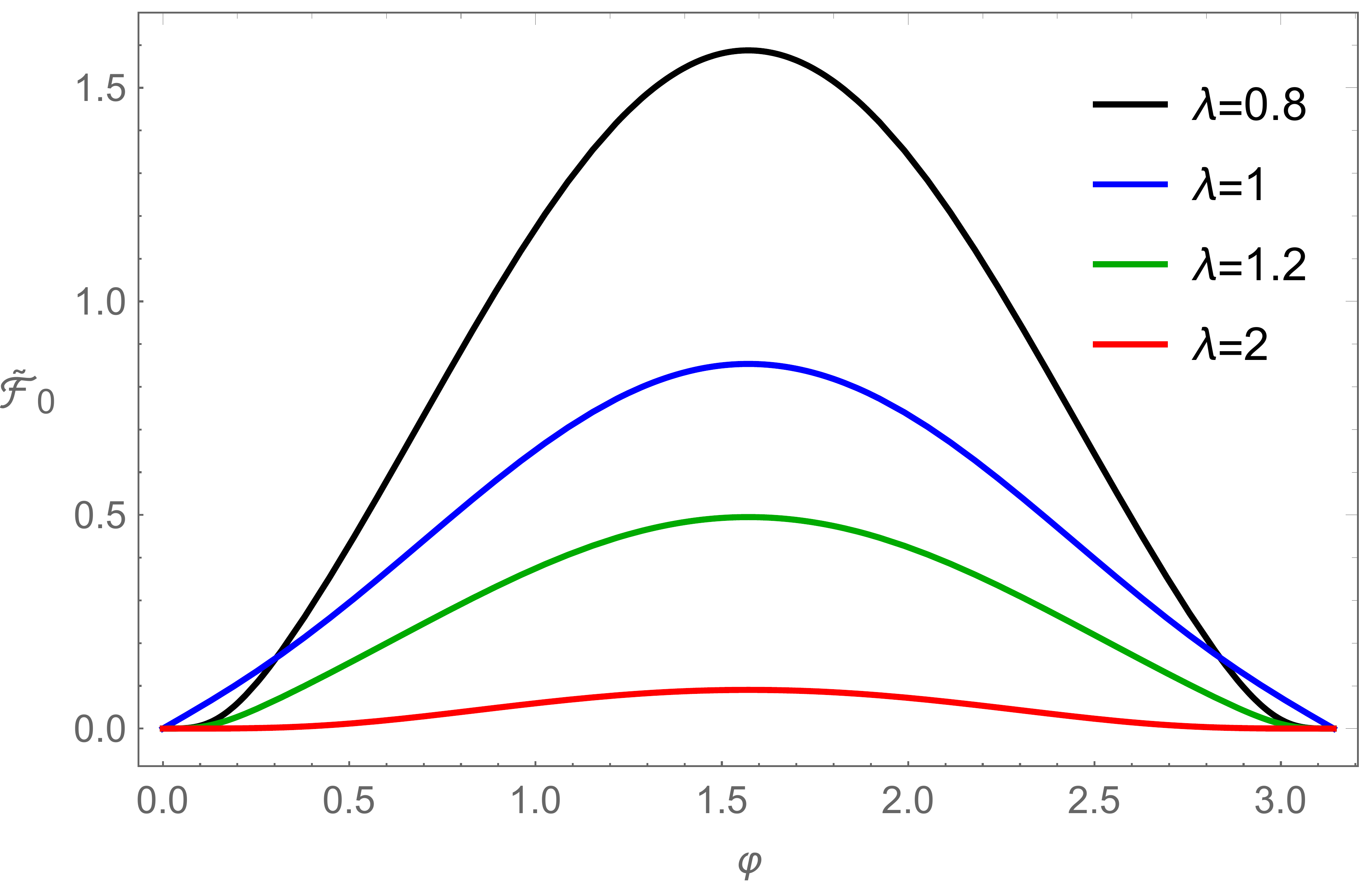}
		\caption{The dependence of $\tilde{F}_{0}(\lambda,\varphi)$ of $\lambda$ with fixed values of $\varphi$ (left pane), and of $\varphi$ with fixed values of $\lambda$ (right pane).}
		\label{fig:F0}
	\end{minipage}
\end{figure}

\begin{figure}[h]
	\begin{minipage}{\textwidth}
		\centering
		\includegraphics[width=0.48\textwidth]{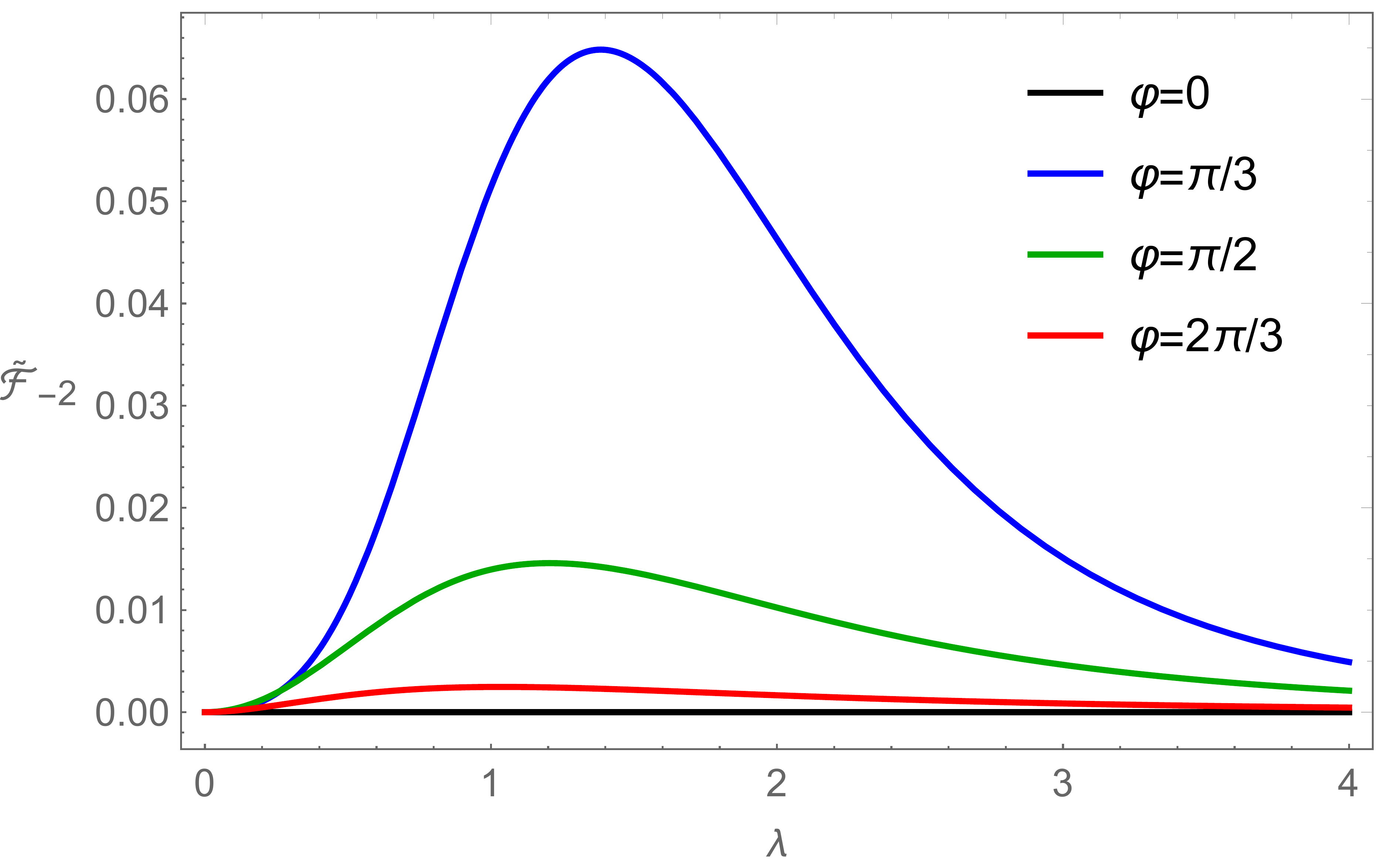}
		\includegraphics[width=0.48\textwidth]{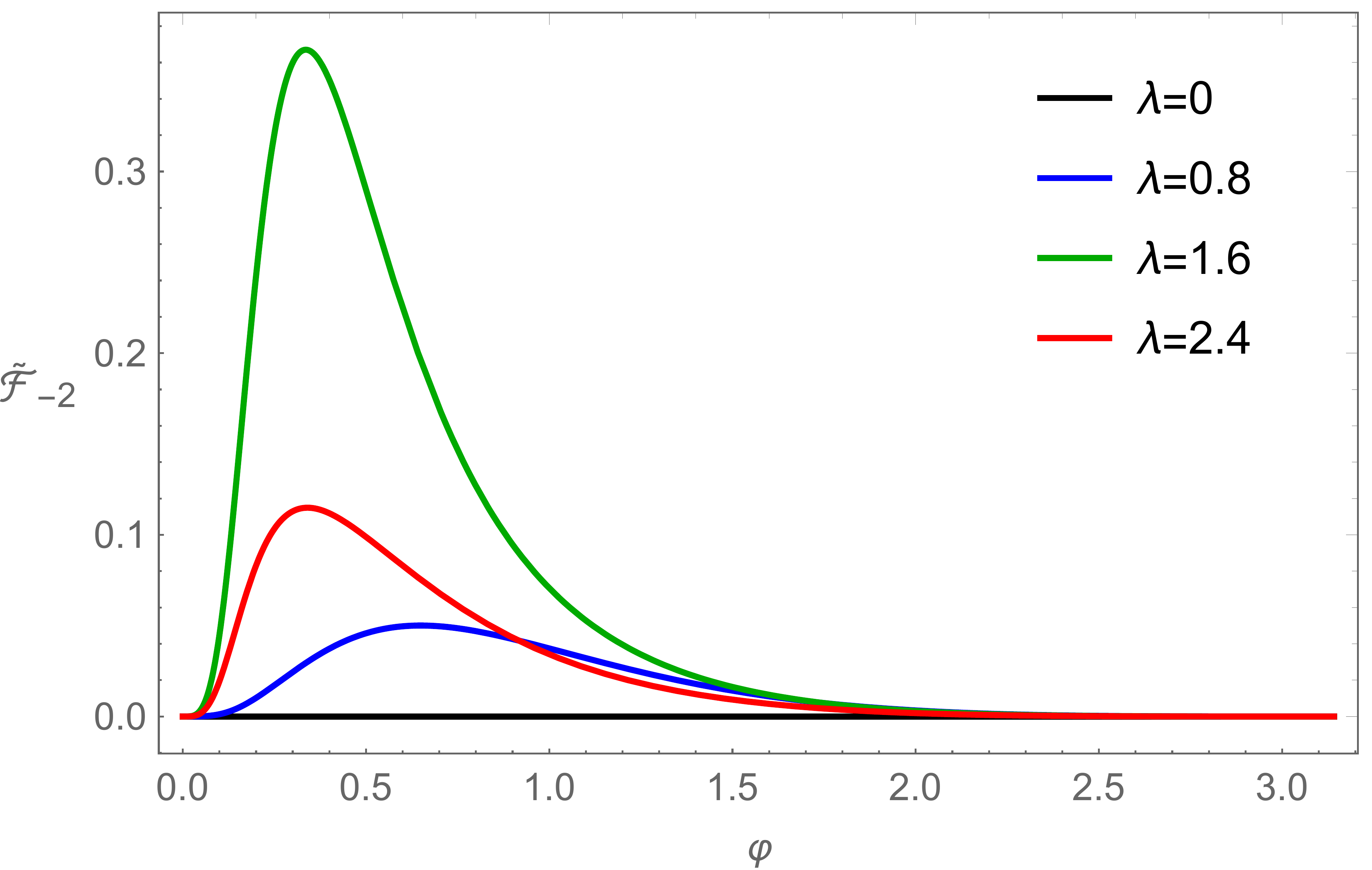}
		\caption{The dependence of $\tilde{F}_{-2}(\lambda,\varphi)$ of $\lambda$ with fixed values of $\varphi$ (left pane), and of $\varphi$ with fixed values of $\lambda$ (right pane).}
		\label{fig:Fm2}
	\end{minipage}
\end{figure}

Without loss of generality, we assume $\bm{k}=k\left(0,0,1\right)$ and $\bm{k}_{\ast}=k_{\ast}\left(\sin\varphi,0,\cos\varphi\right)$. With this convention, the circular polarization tensors in (\ref{tildecalF_p2})-(\ref{tildecalF_m2}) can be evaluated explicitly. In particular, $\tilde{\mathcal{F}}_{0,\pm 2}$ are functions of the angle $\varphi$ between $\bm{k}$ and $\bm{k}_{\ast}$ as well as the ratio
	\begin{equation}
		\lambda = k/k_{\ast},
	\end{equation}
only, which are given by
	\begin{eqnarray}
	\tilde{\mathcal{F}}_{2}=\tilde{\mathcal{F}}_{2}\left(\lambda,\varphi\right) & = & \frac{1}{2}\frac{k_{\ast}^{4}k^{5}\sin^{4}\varphi}{k^{2}+4kk_{\ast}\cos\varphi+4k_{\ast}^{2}}\mathcal{F}\left(k,\left|\bm{k}+2\bm{k}_{\ast}\right|,\left|\bm{k}+\bm{k}_{\ast}\right|\right),\\
	\tilde{\mathcal{F}}_{0}=\tilde{\mathcal{F}}_{0}\left(\lambda,\varphi\right) & = & \frac{1}{2}k^{3}k_{\ast}^{4}\sin^{4}\varphi\left(\mathcal{F}\left(k,k,\left|\bm{k}+\bm{k}_{\ast}\right|\right)+\mathcal{F}\left(k,k,\left|\bm{k}-\bm{k}_{\ast}\right|\right)\right),\\
	\tilde{\mathcal{F}}_{-2}=\tilde{\mathcal{F}}_{-2}\left(\lambda,\varphi\right) & = & \frac{1}{2}\frac{k_{\ast}^{4}k^{5}\sin^{4}\varphi}{k^{2}-4kk_{\ast}\cos\varphi+4k_{\ast}^{2}}\mathcal{F}\left(k,\left|\bm{k}-2\bm{k}_{\ast}\right|,\left|\bm{k}-\bm{k}_{\ast}\right|\right),
	\end{eqnarray}
where $\mathcal{F}$ is defined in (\ref{calF}).
Note  $\tilde{\mathcal{F}}_{0,\pm 2}$ do not depend on the helicity $s$ any more, that is $\Delta_{++}^{(2)}=\Delta_{--}^{(2)}=\Delta_{+-}^{(2)}=\Delta_{-+}^{(2)}$, which implies that the spectra of the two circular polarization modes as well as their cross-spectra have the same amplitude.
The dependence of $\tilde{\mathcal{F}}_{0,\pm 2}$ with respect to $\lambda$ and $\varphi$ are illustrated in Fig.\ref{fig:Fp2}, Fig.\ref{fig:F0} and Fig.\ref{fig:Fm2}, respectively. 
One may check that
	\begin{eqnarray}
	\tilde{\mathcal{F}}_{2} & \rightarrow & \infty,\qquad\text{for}\qquad \lambda\rightarrow2,\varphi\rightarrow\pi,\\
	\tilde{\mathcal{F}}_{0} & \rightarrow & \infty,\qquad\text{for}\qquad\lambda\rightarrow0,\\
	\tilde{\mathcal{F}}_{-2} & \rightarrow & \infty,\qquad\text{for}\qquad\lambda\rightarrow2,\varphi\rightarrow0.
	\end{eqnarray}
These apparent divergences, however, are somewhat marginal. For example, for contribution proportional to $\tilde{\mathcal{F}}_{2}$, due to the presence of $\delta^{3}\left(\bm{k}+\bm{k}'+2\bm{k}_{\ast}\right)$ in (\ref{Delta_2_ssp_sc}), the limit $\lambda\rightarrow2,\varphi\rightarrow\pi$ implies $\bm{k} + 2\bm{k}_{\ast} = 0$, and thus this divergence is relevant only when $\bm{k}' \rightarrow 0$. 
Same analysis will show that the apparent divergence in $\tilde{\mathcal{F}}_{0}$ is relevant only when both $\bm{k}\rightarrow 0, \bm{k}'\rightarrow0$, and in   $\tilde{\mathcal{F}}_{-2}$ is relevant only when $\bm{k}'\rightarrow 0$.

\section{Scalar perturbation} \label{sec:sp}

The analysis for the scalar perturbation $\zeta$ is completely parallel to that of the tensor perturbations.
Using (\ref{zeta_dec}),  the ``free'' two-point function for $\zeta$ is
	\begin{equation}
	\left\langle \hat{\zeta}\left(\tau,\bm{k}\right)\hat{\zeta}\left(\tau',\bm{k}'\right)\right\rangle^{(0)} =\left(2\pi\right)^{3}\delta^{3}\left(\bm{k}+\bm{k}'\right)\frac{1}{2\epsilon}\frac{v\left(\tau,k\right)}{a\left(\tau\right)}\frac{v^{\ast}\left(\tau',k\right)}{a\left(\tau'\right)},\label{2pf_zz}
	\end{equation}
where the mode function $v(\tau,k)$ is given in (\ref{ms_vw}).
The standard isotropic and homogeneous power spectrum for $\zeta$ is
	\begin{equation}
	\Delta_{\zeta}^{(0)}(\bm{k},\bm{k}')\equiv\left\langle \hat{\zeta}(\tau,\bm{k})\hat{\zeta}(\tau,\bm{k}')\right\rangle ^{(0)}=\left(2\pi\right)^{3}\delta^{3}\left(\bm{k}+\bm{k}'\right)\mathcal{P}_{\zeta}^{(0)}(k), \label{Delta_zeta_0}
	\end{equation}
with
	\begin{equation}
	\mathcal{P}_{\zeta}^{(0)}(k)=\frac{1}{2\epsilon}\left.\frac{\left|v\left(\tau,k\right)\right|^{2}}{a^{2}\left(\tau\right)}\right|_{\tau\rightarrow0}.
	\end{equation}

\begin{figure}[h]
		\begin{centering}
			\includegraphics[scale=0.6]{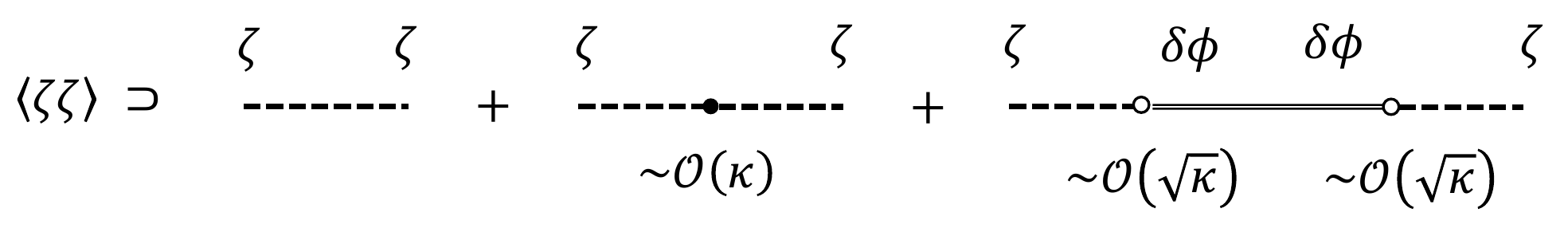}
			\par\end{centering}
		\caption{Illustration of contributions to the power spectrum of the scalar perturbation $\zeta$ up to the linear order in $\kappa$. The total power spectra consist of three contributions: the standard isotropic and homogeneous spectra (left), corrections due to the inhomogeneous background metric $\bar{h}_{ij}$ (middle), corrections due to the couplings with scalar field perturbations $\delta\phi^{I}$ (right). Both corrections are of order $\kappa$.}
		\label{fig:zeta}
	\end{figure}

According to the  in-in formalism, up to the linear order in $\kappa$, the  power spectrum of $\zeta$ is given by
	\begin{equation}
	\left\langle \hat{\zeta}(\bm{k})\hat{\zeta}(\bm{k}')\right\rangle =\left[\Delta_{\zeta}^{(0)}+\Delta_{\zeta}^{(1)}+\Delta_{\zeta}^{(2)}\right](\bm{k},\bm{k}'),
	\end{equation}
where $\Delta_{\zeta}^{(0)}(\bm{k},\bm{k}')$ is the isotropic and homogeneous power spectrum given in (\ref{Delta_zeta_0}), $\Delta_{\zeta}^{(1)}(\bm{k},\bm{k}')$ and $\Delta_{\zeta}^{(2)}(\bm{k},\bm{k}')$ are corrections due to the background inhomogeneities, which are given by
	\begin{equation}
	\Delta_{\zeta}^{(1)}(\bm{k},\bm{k}')=2\Im\left(\int_{-\infty}^{0}\!\mathrm{d}\tau\left\langle \hat{\zeta}(0,\bm{k})\hat{\zeta}(0,\bm{k}')H_{1}^{(\zeta\zeta)}(\tau)\right\rangle \right),
	\end{equation}
and
	\begin{eqnarray}
	\Delta_{\zeta}^{(2)}(\bm{k},\bm{k}') & = & \int_{-\infty}^{0}\!\mathrm{d}\tau_{1}\int_{-\infty}^{0}\!\mathrm{d}\tau_{2}\left\langle H_{1}^{(\zeta\delta\phi)}(\tau_{1})\hat{\zeta}(0,\bm{k})\hat{\zeta}(0,\bm{k}')H_{1}^{(\zeta\delta\phi)}(\tau_{2})\right\rangle \nonumber \\
	&  & -2\Re\left(\int_{-\infty}^{0}\!\mathrm{d}\tau_{1}\int_{-\infty}^{\tau_{1}}\!\mathrm{d}\tau_{2}\left\langle \hat{\zeta}(0,\bm{k})\hat{\zeta}(0,\bm{k}')H_{1}^{(\zeta\delta\phi)}(\tau_{1})H_{1}^{(\zeta\delta\phi)}(\tau_{2})\right\rangle \right),
	\end{eqnarray}
where $H_{1}^{(\zeta\zeta)}$ and $H_{1}^{(\zeta\delta\phi)}$ are given in (\ref{Ham_zz_1}) and (\ref{Ham_zf_1}), respectively.
The two types of corrections are depicted in the Feynman-type diagrams in Fig.\ref{fig:zeta}.
In particular, $\Delta_{\zeta}^{(1)}(\bm{k},\bm{k}')$ is the correction from the inhomogeneous background metric $\bar{h}_{ij} = \delta_{ij}+X_{ij}$ directly, $\Delta_{\zeta}^{(2)}(\bm{k},\bm{k}')$ is the correction from the couplings between  $\zeta$ and perturbations of the scalar fields $\delta\phi^{I}$, of which the existence is due to the background inhomogeneity.

After some manipulations, we find
	\begin{equation}
	\Delta_{\zeta}^{(1)}(\bm{k},\bm{k}')=-\frac{H^{2}}{\epsilon}\frac{1}{kk'\left(k+k'\right)}\left[\mathcal{C}(-\bm{k},-\bm{k}',\bm{k}+\bm{k}')+\frac{1}{4}\mathcal{D}(-\bm{k},-\bm{k}',\bm{k}+\bm{k}')\frac{k^{2}+k'^{2}+k'k}{k'^{2}k^{2}}\right], \label{Delta_1_zeta_fin}
	\end{equation}
and
	\begin{eqnarray}
	&  & \Delta_{\zeta}^{(2)}\left(\bm{k},\bm{k}'\right)\nonumber \\
	& = & H^{2}\int\frac{\mathrm{d}^{3}p}{\left(2\pi\right)^{3}}\bigg\{\frac{(k'+k+p)}{4k'kp(k'+k)(k'+p)(k+p)}\bigg[2p^{2}\mathcal{C}^{I}\left(-\bm{k},-\bm{p},\bm{k}+\bm{p}\right)\mathcal{C}^{I}\left(-\bm{k}',\bm{p},\bm{k}'-\bm{p}\right)\nonumber \\
	&  & \qquad+\mathcal{C}^{I}\left(-\bm{k},-\bm{p},\bm{k}+\bm{p}\right)\mathcal{D}^{I}\left(-\bm{k}',\bm{p},\bm{k}'-\bm{p}\right)+\mathcal{C}^{I}\left(-\bm{k}',\bm{p},\bm{k}'-\bm{p}\right)\mathcal{D}^{I}\left(-\bm{k},-\bm{p},\bm{k}+\bm{p}\right)\bigg]\nonumber \\
	&  & \quad+\mathcal{F}\left(k,k',p\right)\mathcal{D}^{I}\left(-\bm{k},-\bm{p},\bm{k}+\bm{p}\right)\mathcal{D}^{I}\left(-\bm{k}',\bm{p},\bm{k}'-\bm{p}\right)\bigg\}. \label{Delta_2_zeta_fin}
	\end{eqnarray}
where $\mathcal{F}$ is defined by (\ref{calF}).
At this point, since we assume $\sqrt{\mathcal{C}}\sim \sqrt{\mathcal{D}}\sim\mathcal{C}^I\sim\mathcal{D}^I \sim \mathcal{O}(\sqrt{\kappa})$,
	\begin{eqnarray}
		\Delta_{\zeta}^{(2)} \sim \epsilon \, \Delta_{\zeta}^{(1)},
	\end{eqnarray}
which implies the corrections to the power spectrum of $\zeta$ are dominated by $\Delta_{\zeta}^{(1)}$, i.e., the contribution from the self-coupling of $\zeta$ due to the background inhomogeneities.

\subsection{Contributions from background inhomogeneities}

In the following, we evaluate the contributions to $\Delta_{\zeta}^{(1)}(\bm{k},\bm{k}')$ from $\alpha$, $\beta$, $\xi_{i}$ and $\eta_{ij}$ defined in (\ref{Xij_dec}) separately, which are parallel to those performed in Sec.\ref{sec:cbi_gw}.

\subsubsection{Contribution from $\alpha$}

Plugging the expressions for $\mathcal{C}$ and $\mathcal{D}$ (\ref{calC_fin})-(\ref{calD_fin}) into (\ref{Delta_1_zeta_fin}) and using  $X_{ij}^{(\alpha)} = 2\alpha \delta_{ij}$,
we get the contribution to $\Delta_{\zeta}^{(1)}$ from $X_{ij}^{(\alpha)} = 2\alpha \delta_{ij}$ 
	\begin{equation}
	\Delta_{\zeta}^{(1),\alpha}\left(\bm{k},\bm{k}'\right)=\frac{H^{2}}{\epsilon\, k^{3}}\frac{3r+\left(1+r+r^{2}\right)\cos\theta}{4r^{2}\left(1+r\right)}\alpha\left(\bm{k}+\bm{k}'\right),
	\end{equation}
where $r=k'/k$ and $\cos\theta = \hat{\bm{k}}\cdot \hat{\bm{k}}'$.

\subsubsection{Contribution from $\beta$}

The contribution to $\Delta_{\zeta}^{(1)}$ from $X_{ij}^{(\beta)} = 2\left(\partial_{i}\partial_{j}-\frac{1}{3}\delta_{ij}\partial^{2}\right)\beta$ is given by
\begin{equation}
\Delta_{\zeta}^{(1),\beta}\left(\bm{k},\bm{k}'\right)=\frac{H^{2}}{\epsilon\,k}\frac{1+r^{2}+r}{6r^{2}\left(1+r\right)}\left[3r+2\left(r^{2}+1\right)\cos\theta+r\cos^{2}\theta\right]\beta\left(\bm{k}+\bm{k}'\right)
\end{equation}

\subsubsection{Contribution from $\xi_{i}$}

Under the same decomposition for $\xi_{i}$ as in (\ref{xi_dec}), the contribution to $\Delta_{\zeta}^{(1)}$ from $X_{ij}^{(\xi)} = \partial_{i}\xi_{j}+\partial_{j}\xi_{i}$ is
	\begin{equation}
	\Delta_{\zeta}^{(1),\xi}\left(\bm{k},\bm{k}'\right)=\frac{H^{2}}{\epsilon\,k^{2}}\frac{\left(1-r^{3}\right)\sin\theta}{4\sqrt{2}r^{2}\sqrt{1+2r\cos\theta+r^{2}}}x_{+}\left(\bm{k}+\bm{k}'\right),
	\end{equation}
where $x_{+}$ is defined in (\ref{xpm_def}).

\subsubsection{Contribution from $\eta_{ij}$}

Under the same decomposition for $\eta_{ij}$ as in (\ref{eta_dec}), 
the contribution to $\Delta_{\zeta}^{(1)}$ from $X_{ij}^{(\eta)} = \eta_{ij}$ is
\begin{eqnarray}
\Delta_{\zeta}^{(1),\eta}\left(\bm{k},\bm{k}'\right) & = & -\frac{H^{2}}{16\epsilon\,k^{3}}\frac{1+r^{2}+r}{r^{2}\left(1+r\right)\left(1+2r\cos\theta+r^{2}\right)}\nonumber \\
&  & \times\left[r+\left(2+r^{2}\right)\cos\theta+3r\cos(2\theta)+r^{2}\cos(3\theta)\right]y_{+}\left(\bm{k}+\bm{k}'\right),
\end{eqnarray}
where $y_{+}$ is defined in (\ref{ypm_def}).

As a consistency check, all the above contributions are vanishing in the homogeneous limit, 
	\begin{equation}
		\alpha, \beta , x_{+} , y_{+} \rightarrow \delta^3(\bm{k}+\bm{k}'),
	\end{equation}
which corresponds to $r \rightarrow 1$ and $\theta\rightarrow \pi$.

\subsection{Contributions from interactions with $\delta\phi^{I}$}

Finally we evaluate $\Delta_{\zeta}^{(2)}$ for completeness.
Plugging the expressions for $\mathcal{C}^I{I}$ and $\mathcal{D}^{I}$ in (\ref{calCIDI_fin}) into (\ref{Delta_2_zeta_fin}), we get
	\begin{equation}
	\Delta_{\zeta}^{(2)}\left(\bm{k},\bm{k}'\right)=H^{2}\int\frac{\mathrm{d}^{3}p}{\left(2\pi\right)^{3}}\left(k^{2}+\bm{k}\cdot\bm{p}\right)\left(k'^{2}-\bm{k}'\cdot\bm{p}\right)\bar{\phi}^{I}\left(\bm{k}+\bm{p}\right)\bar{\phi}^{I}\left(\bm{k}'-\bm{p}\right)\mathcal{G}\left(k,k',p\right), \label{Delta_2_zeta_xpl}
	\end{equation}
with
	\begin{eqnarray}
	\mathcal{G}\left(k,k',p\right) & \equiv & \frac{1}{8k'^{3}k^{3}p^{3}(k'+k)(k'+p)(k+p)}\Big[p^{5}+p^{4}(k'+k)\nonumber \\
	&  & +p^{3}\left(7k'^{2}+8k'k+7k^{2}\right)+p^{2}(k'+k)\left(7k'^{2}+8k'k+7k^{2}\right)\nonumber \\
	&  & +8pk'k(k'+k)^{2}+8k'^{2}k^{2}(k'+k)\Big].
	\end{eqnarray}
Giving the configuration for the background values of $\bar{\phi}^I$ (such as the ansatz in (\ref{phibar_ansatz})), one is able to evaluate the integral (\ref{Delta_2_zeta_xpl}) explicitly.
As we have discussed below (\ref{Delta_2_zeta_fin}), however, the dominant correction to the power spectrum of $\zeta$ is $\Delta_{\zeta}^{(1)}$ and thus we do not evaluate $\Delta_{\zeta}^{(2)}$ explicitly.

\section{Conclusion} \label{sec:con}

In this work we have studied inhomogeneous inflation scenario, i.e., exponential expansion of the space with inhomogeneities, and its cosmological perturbations.
Such an inhomogeneous inflation can be realized in a simple model of multiple scalar fields minimally coupled to gravity.
One of these scalar fields is the usual inflaton field $\Phi$, which is timelike and can be chosen to be spatially homogeneous.
Other scalar fields $\phi^{I}$  have spacelike gradients and thus are always spatially inhomogeneous.
In particular, at the background level we show that our model possesses a solution in which the background values of $\phi^{I}$ is time-independent $\bar{\phi}^{I} = \bar{\phi}^{I}(\vec{x})$, and the space uniformly expand in different directions with the same rate $\bar{g}_{ij}(\tau,\vec{x}) = a^2(\tau)\bar{h}_{ij}(\vec{x})$.

We have also investigated cosmological perturbations around such an inhomogeneous inflation background.
We assume the level of inhomogeneities of the background is small, which is of order $\kappa$. 
By using the perturbative approach, we calculated the corrections to the power spectra of gravitational waves and curvature perturbation up to the linear order in $\kappa$.
Since both homogeneity and isotropy get lost with our background, perturbation modes with different wave numbers $k$ and $k'$ are correlated, and the correlations depend on both the ratio $r=k'/k$ and the angle $\cos\theta = \hat{\bm{k}}\cdot\hat{\bm{k}}'$.
In particular, we find the two circular polarization modes of gravitational waves are correlated generally, and will have the same power when the background inhomogeneities are of ``scalar-type'', i.e. $X_{ij} \supset 2\alpha\delta_{ij}$ or $X_{ij} \supset \left(\partial_{i}\partial_{j}-\frac{1}{3}\delta_{ij}\partial^{2}\right)\beta$, and will have different powers if the background inhomogeneities are of ``vector-type'' $X_{ij} \supset \partial_{i}\xi_{j}+\partial_{j}\xi_{i}$ or ``tensor-type'' $X_{ij} \supset \eta_{ij}$.

As a final remark, note linear perturbations on an inhomogeneous background are tightly related to non-linear perturbations on a homogeneous background. 
Thus it will be useful to put constraints on the deviation of homogeneity by using the current observation of primordial non-Gaussianities.


\acknowledgments

This work was supported by the Chinese National
Youth Thousand Talents Program (No. 71000-41180003) and by the SYSU start-up funding (No. 71000-52601106).

\appendix

\section{Gauge issues}\label{sec:gauge}

Under infinitesimal coordinate transformation $x^{\mu} \rightarrow x^{\mu} + \xi{^\mu}$, the linear gauge transformation of the perturbation of a spacetime tensor field $Q$ around some background value $\bar{Q}$ is given by $\Delta_{\bm{\xi}}\left(\delta Q\right)=-\pounds_{\bm{\xi}}\bar{Q}$.
In our case, since $\bar{\Phi} = \bar{\Phi}(\tau)$ and $\bar{\phi}^I= \bar{\phi}^I(\vec{x})$,
\begin{equation}
\delta\Phi \rightarrow \delta\Phi-\xi^{0}\bar{\Phi}',\qquad \delta\phi^{I} \rightarrow \delta\phi^{I} -\xi^{i}\partial_{i}\bar{\phi}^{I}.
\end{equation}
We may write $\xi^{i}\equiv\bar{h}^{ij}\left(\partial_{j}\xi_{\mathrm{L}}+\tilde{\xi}_{j}\right)$
with $\bar{\nabla}^i \tilde{\xi}_i =0$. For the metric perturbations, straightforward algebra shows
\begin{eqnarray}
A & \rightarrow & A-\frac{1}{a}\left(a\xi^{0}\right)',\\
B & \rightarrow & B+\xi^{0}-\xi_{\mathrm{L}}',\\
\tilde{B}_{i} & \rightarrow & \tilde{B}_{i}-\tilde{\xi}_{i}',\\
\zeta & \rightarrow & \zeta-\frac{a'}{a}\xi^{0}-\frac{1}{3}\bar{\nabla}^{2}\xi_{\mathrm{L}},\\
E & \rightarrow & E-\xi_{\mathrm{L}},\\
F_{i} & \rightarrow & F_{i}-\tilde{\xi}_{i},\\
\gamma_{ij} & \rightarrow & \gamma_{ij}.
\end{eqnarray}

From the above, we find that the following combination
\begin{equation}
\hat{\zeta}:=\zeta-\frac{1}{3}\bar{\nabla}^{2}E-\frac{\mathcal{H}}{\bar{\Phi}'}\delta\Phi,
\end{equation}
is gauge invariant.
In practise, we may completely fix the gauge freedom by setting  $\delta\Phi = E=F_i =0$ or $\zeta = E = F_i =0$. 
Note in the case of ``solid inflation'' with $I=1,2,3$, one may also choose the so-called ``unitary gauge'' by setting $\delta\Phi = \delta\phi^{I} = 0$.

\section{Details in deriving the quadratic Lagrangian} \label{sec:details}

Straightforward expansion of the Lagrangian (\ref{model}) around the inhomogeneous background up to the quadratic order in $A$, $B_i$ and $H_{ij}$ yields
	\begin{eqnarray}
	\mathcal{L}_{2} & = & -a^{2}\mathcal{H}^{2}\left(3-\tilde{\epsilon}\right)A^{2}+a^{2}\bigg(\mathcal{H}H'^{i}{}_{i}-\frac{1}{2}\mathcal{K}H^{i}{}_{i}+\frac{1}{2}\bar{\mathrm{D}}_{i}\bar{\mathrm{D}}_{j}H^{ij}-\frac{1}{2}\bar{\mathrm{D}}^{2}H^{i}{}_{i}-\bar{\Phi}'\delta\Phi'-a^{2}\bar{V}_{,\Phi}\delta\Phi\nonumber \\
	&  & -\bar{\mathrm{D}}^{i}\left(\bar{f}_{IJ}\bar{\mathrm{D}}_{i}\bar{\phi}^{I}\delta\phi^{J}\right)\bigg)A-a^{2}\left(\frac{1}{2}\bar{\mathrm{D}}_{i}H'^{j}{}_{j}-\frac{1}{2}\bar{\mathrm{D}}^{j}H'_{ij}+\bar{\Phi}'\bar{\mathrm{D}}_{i}\delta\Phi+\bar{f}_{IJ}\bar{\mathrm{D}}_{i}\bar{\phi}^{J}\delta\phi'^{I}\right)B^{i}\nonumber \\
	&  & -2a^{2}\mathcal{H}A\bar{\mathrm{D}}_{i}B^{i}+\frac{a^{2}}{4}\bar{\mathrm{D}}_{j}B_{i}\bar{\mathrm{D}}^{j}B^{i}-\frac{a^{2}}{4}\bar{\mathrm{D}}_{i}B^{i}\bar{\mathrm{D}}_{j}B^{j}+\frac{a^{2}}{4}\left(\bar{\mathcal{R}}_{ij}-2\mathcal{K}\bar{h}_{ij}\right)B^{i}B^{j}\nonumber \\
	&  & +\frac{a^{2}}{8}\left(H'_{ij}H'^{ij}-H'^{i}{}_{i}H'^{j}{}_{j}\right)+\frac{a^{2}}{8}\mathcal{K}\left(2H^{ij}H_{ij}-H^{i}{}_{i}H^{j}{}_{j}\right)-\frac{a^{2}}{8}H^{i}{}_{i}\bar{\mathrm{D}}^{2}H^{j}{}_{j}\nonumber \\
	&  & +\frac{a^{2}}{4}H^{i}{}_{i}\bar{\mathrm{D}}_{j}\bar{\mathrm{D}}_{k}H^{jk}-\frac{a^{2}}{4}H^{ij}\bar{\mathrm{D}}_{k}\bar{\mathrm{D}}_{j}H_{i}{}^{k}+\frac{a^{2}}{8}H_{ij}\bar{\mathrm{D}}^{2}H^{ij}+\frac{a^{2}}{2}\delta\Phi'^{2}-\frac{a^{2}}{2}\bar{\mathrm{D}}_{i}\delta\Phi\bar{\mathrm{D}}^{i}\delta\Phi\nonumber \\
	&  & -\frac{a^{4}}{2}\bar{V}_{,\Phi\Phi}\delta\Phi^{2}+\frac{a^{2}}{2}\bar{\Phi}'H^{i}{}_{i}\delta\Phi'-\frac{a^{4}}{2}\bar{V}_{,\Phi}H^{i}{}_{i}\delta\Phi+\frac{a^{2}}{2}\bar{f}_{IJ}\delta\phi'^{I}\delta\phi'^{J}-\frac{a^{2}}{2}\bar{f}_{IJ}\bar{\mathrm{D}}_{i}\delta\phi^{I}\bar{\mathrm{D}}^{i}\delta\phi^{J}\nonumber \\
	&  & -a^{2}\bar{h}^{ij}\bar{\mathrm{D}}_{i}\bar{\phi}^{I}\partial_{K}\bar{f}_{IJ}\delta\phi^{K}\bar{\mathrm{D}}_{j}\delta\phi^{J}-\frac{a^{2}}{4}\bar{\mathrm{D}}_{i}\bar{\phi}^{I}\bar{\mathrm{D}}^{i}\bar{\phi}^{J}\partial_{K}\partial_{L}\bar{f}_{IJ}\delta\phi^{K}\delta\phi^{L}\nonumber \\
	&  & +a^{2}\left(H^{ij}-\frac{1}{2}H^{k}{}_{k}\bar{h}^{ij}\right)\left(\bar{f}_{IJ}\bar{\mathrm{D}}_{i}\bar{\phi}^{I}\bar{\mathrm{D}}_{j}\delta\phi^{J}+\frac{1}{2}\bar{\mathrm{D}}_{i}\bar{\phi}^{I}\bar{\mathrm{D}}_{j}\bar{\phi}^{J}\partial_{K}\bar{f}_{IJ}\delta\phi^{K}\right),\label{L2_ori}
	\end{eqnarray}
where $\bar{\mathrm{D}}^{i}\equiv\bar{h}^{ij}\bar{\mathrm{D}}_{j}$. 
In deriving (\ref{L2_ori}) we have already used background equations of motion to simplify some coefficients. At this point, we have not chosen any gauge and thus (\ref{L2_ori}) is exact.

In (\ref{L2_ori}), $A$ and $B_i\equiv \partial_{i} B + \tilde{B}_i$ contain no time derivatives and thus play as auxiliary variables.
Varying (\ref{L2_ori}) with respect to  $A$, $B$ and $\tilde{B}_i$ yields the following constraint equations
	\begin{eqnarray}
	2\mathcal{H}^{2}\left(3-\tilde{\epsilon}\right)A+2\mathcal{H}\bar{\mathrm{D}}^{2}B-\mathcal{C} & = & 0,\label{eom_A}\\
	2\mathcal{H}\bar{\mathrm{D}}^{2}A-\mathcal{K}\bar{\mathrm{D}}^{2}B-\bar{\mathrm{D}}^{2}\mathcal{D} & = & 0,\label{eom_B}\\
	\frac{1}{2}\left(\bar{h}_{ij}\bar{\mathrm{D}}^{2}-\bar{\mathcal{R}}_{ij}+2\mathcal{K}\bar{h}_{ij}\right)\tilde{B}^{j}+\tilde{\mathcal{D}}_{i} & = & 0,\label{eom_Bti}
	\end{eqnarray}
where $\tilde{\epsilon}$ is defined in (\ref{epsilon_def}),
\begin{eqnarray}
\mathcal{C} & \equiv & \mathcal{H}H'^{i}{}_{i}-\frac{1}{2}\mathcal{K}H^{i}{}_{i}+\frac{1}{2}\bar{\mathrm{D}}_{i}\bar{\mathrm{D}}_{j}H^{ij}-\frac{1}{2}\bar{\mathrm{D}}^{2}H^{i}{}_{i}-\bar{\Phi}'\delta\Phi'\nonumber \\
&  & -a^{2}\bar{V}_{,\Phi}\delta\Phi-\bar{\mathrm{D}}^{i}\left(\bar{f}_{IJ}\bar{\mathrm{D}}_{i}\bar{\phi}^{I}\delta\phi^{J}\right),\label{calC_def}
\end{eqnarray}
$\mathcal{D}$ and $\tilde{\mathcal{D}}_i$ are irreducible parts of $\mathcal{D}_{i}$ given by
\begin{eqnarray} 
\mathcal{D}_{i} & \equiv & \frac{1}{2}\bar{\mathrm{D}}_{i}H'^{j}{}_{j}-\frac{1}{2}\bar{\mathrm{D}}^{j}H'_{ij}+\bar{\Phi}'\bar{\mathrm{D}}_{i}\delta\Phi+\bar{f}_{IJ}\bar{\mathrm{D}}_{i}\bar{\phi}^{J}\delta\phi'^{I},\label{calDi_def}
\end{eqnarray}
with $\mathcal{D}_i \equiv \bar{\mathrm{D}}_{i}\mathcal{D}+\tilde{\mathcal{D}}_{i}$ and $\bar{\mathrm{D}}_{i}\tilde{\mathcal{D}}^{i}=0$.
From (\ref{eom_A})-(\ref{eom_Bti}) we can solve $A$, $B$ and $\tilde{B}_i$ in terms of $H_{ij}$, $\delta\Phi$ and $\delta\phi^I$ to be
	\begin{eqnarray}
	A & = & \left(\bar{\mathrm{D}}^{2}+\frac{3-\tilde{\epsilon}}{2}\mathcal{K}\right)^{-1}\left(\frac{\mathcal{K}}{4\mathcal{H}^{2}}\mathcal{C}+\frac{1}{2\mathcal{H}}\bar{\mathrm{D}}^{2}\mathcal{D}\right),\label{sol_A}\\
	B & = & \left(\bar{\mathrm{D}}^{2}+\frac{3-\tilde{\epsilon}}{2}\mathcal{K}\right)^{-1}\left(\frac{1}{2\mathcal{H}}\mathcal{C}-\frac{3-\tilde{\epsilon}}{2}\mathcal{D}\right),\label{sol_B}\\
	\tilde{B}^{i} & = & -2\left(\Xi^{-1}\right)^{ij}\tilde{\mathcal{D}}_{j},\label{sol_Bti}
	\end{eqnarray}
where $\left(\Xi^{-1}\right)^{ij}$ is the formal inverse of 
\begin{equation}
\Xi_{ij}\equiv\bar{h}_{ij}\bar{\mathrm{D}}^{2}-\bar{\mathcal{R}}_{ij}+2\mathcal{K}\bar{h}_{ij}.\label{Xi_def}
\end{equation}
Note (\ref{sol_A})-(\ref{sol_Bti}) can be compared with the corresponding results in usual cosmological perturbations in FRW background (e.g. \cite{Gao:2011qe}) or solid inflation \cite{Endlich:2012pz}.

Plugging (\ref{sol_A})-(\ref{sol_Bti}) into (\ref{L2_ori}) and by choosing the gauge $E=F_i=\delta\Phi=0$, after tedious calculations, we arrive at the full quadratic Lagrangian for perturbation variables $\zeta$, $\gamma_{ij}$, and $\delta^I$ in (\ref{L2_fin}), in which various terms are given by
\begin{eqnarray}
\mathcal{L}_{2}^{(\zeta\zeta)} & = & a^{2}\tilde{\epsilon}\,\zeta'^{2}+\frac{1}{2}\mathcal{K}a^{2}\tilde{\epsilon}^{2}\,\zeta'\left(\bar{\mathrm{D}}^{2}+\frac{3-\tilde{\epsilon}}{2}\mathcal{K}\right)^{-1}\zeta'+a^{2}\tilde{\epsilon}\,\zeta\,\bar{\mathrm{D}}^{2}\zeta\nonumber \\
&  & +\frac{1}{2}\mathcal{K}a^{2}\left(4\tilde{\epsilon}+\tilde{\epsilon}^{2}+\frac{\tilde{\epsilon}'}{\mathcal{H}}\right)\zeta^{2}+\frac{1}{2}\mathcal{K}^{2}a^{2}\left[\frac{1}{2}\tilde{\epsilon}^{2}\left(1+\tilde{\epsilon}\right)+\tilde{\epsilon}\frac{\tilde{\epsilon}'}{\mathcal{H}}\right]\zeta\left(\bar{\mathrm{D}}^{2}+\frac{3-\tilde{\epsilon}}{2}\mathcal{K}\right)^{-1}\zeta\nonumber \\
&  & +\frac{1}{8}\mathcal{K}^{3}a^{2}\frac{\tilde{\epsilon}'}{\mathcal{H}}\tilde{\epsilon}^{2}\,\zeta\left(\bar{\mathrm{D}}^{2}+\frac{3-\tilde{\epsilon}}{2}\mathcal{K}\right)^{-2}\zeta,\label{L2_zz}
\end{eqnarray}
and
\begin{equation}
\mathcal{L}_{2}^{(\gamma\gamma)}=\frac{a^{2}}{8}\gamma_{ij}'\gamma'^{ij}+\frac{a^{2}}{8}\gamma^{ij}\bar{\mathrm{D}}^{2}\gamma_{ij}-\frac{a^{2}}{4}\left[3\bar{f}_{ij}-\frac{1}{2}\left(\bar{h}^{kl}\bar{f}_{kl}-\mathcal{K}\right)\bar{h}_{ij}\right]\gamma^{ki}\gamma_{k}{}^{j}.\label{L2_gg}
\end{equation}
and
\begin{eqnarray}
\mathcal{L}_{2}^{(\delta\phi\delta\phi)} & = & \frac{a^{2}}{2}\bar{f}_{IJ}\delta\phi'^{I}\delta\phi'^{J}+a^{2}\Delta_{i}^{\phantom{i}k}\left(\bar{f}_{IJ}\bar{\mathrm{D}}_{k}\bar{\phi}^{J}\delta\phi'^{I}\right)\left(\Xi^{-1}\right)^{ij}\Delta_{j}^{\phantom{i}m}\left(\bar{f}_{IJ}\bar{\mathrm{D}}_{m}\bar{\phi}^{J}\delta\phi'^{I}\right)\nonumber \\
&  & -a^{2}\frac{3-\tilde{\epsilon}}{4}\bar{\mathrm{D}}^{-2}\bar{\mathrm{D}}^{i}\left(\bar{f}_{IJ}\bar{\mathrm{D}}_{i}\bar{\phi}^{J}\delta\phi'^{I}\right)\left(\bar{\mathrm{D}}^{2}+\frac{3-\tilde{\epsilon}}{2}\mathcal{K}\right)^{-1}\bar{\mathrm{D}}^{i}\left(\bar{f}_{IJ}\bar{\mathrm{D}}_{i}\bar{\phi}^{J}\delta\phi'^{I}\right)\nonumber \\
&  & -\frac{a^{2}}{2}\bar{f}_{IJ}\bar{\mathrm{D}}_{i}\delta\phi^{I}\bar{\mathrm{D}}^{i}\delta\phi^{J}-a^{2}\bar{\mathrm{D}}_{i}\bar{\phi}^{I}\partial_{K}\bar{f}_{IJ}\delta\phi^{K}\bar{\mathrm{D}}^{i}\delta\phi^{J}-\frac{a^{2}}{4}\bar{\mathrm{D}}_{i}\bar{\phi}^{I}\bar{\mathrm{D}}^{i}\bar{\phi}^{J}\partial_{K}\partial_{L}\bar{f}_{IJ}\delta\phi^{K}\delta\phi^{L}\nonumber \\
&  & +\frac{1}{4}a^{2}\left(1+\tilde{\epsilon}\right)\bar{\mathrm{D}}^{i}\left(\bar{f}_{IJ}\bar{\mathrm{D}}_{i}\bar{\phi}^{I}\delta\phi^{J}\right)\left(\bar{\mathrm{D}}^{2}+\frac{3-\tilde{\epsilon}}{2}\mathcal{K}\right)^{-1}\bar{\mathrm{D}}^{j}\left(\bar{f}_{KL}\bar{\mathrm{D}}_{j}\bar{\phi}^{K}\delta\phi^{L}\right)\nonumber \\
&  & +\frac{1}{8}\mathcal{K}a^{2}\frac{\tilde{\epsilon}'}{\mathcal{H}}\bar{\mathrm{D}}^{i}\left(\bar{f}_{IJ}\bar{\mathrm{D}}_{i}\bar{\phi}^{I}\delta\phi^{J}\right)\left(\bar{\mathrm{D}}^{2}+\frac{3-\tilde{\epsilon}}{2}\mathcal{K}\right)^{-2}\bar{\mathrm{D}}^{j}\left(\bar{f}_{KL}\bar{\mathrm{D}}_{j}\bar{\phi}^{K}\delta\phi^{L}\right),\label{L2_ff}
\end{eqnarray}
and 
\begin{eqnarray}
\mathcal{L}_{2}^{(\zeta\delta\phi)} & = & a^{2}\tilde{\epsilon}\left(\bar{\mathrm{D}}^{2}+\frac{3-\tilde{\epsilon}}{2}\mathcal{K}\right)^{-1}\zeta'\,\bar{\mathrm{D}}^{i}\left(\bar{f}_{IJ}\bar{\mathrm{D}}_{i}\bar{\phi}^{J}\delta\phi'^{I}\right)+a^{2}\tilde{\epsilon}\,\zeta\,\bar{\mathrm{D}}^{i}\left(\bar{f}_{IJ}\bar{\mathrm{D}}_{i}\bar{\phi}^{I}\delta\phi^{J}\right)\nonumber \\
&  & +\frac{1}{2}\mathcal{K}a^{2}\left[\tilde{\epsilon}\left(1+\tilde{\epsilon}\right)+\frac{\tilde{\epsilon}'}{\mathcal{H}}\right]\left(\bar{\mathrm{D}}^{2}+\frac{3-\tilde{\epsilon}}{2}\mathcal{K}\right)^{-1}\zeta\,\bar{\mathrm{D}}^{i}\left(\bar{f}_{IJ}\bar{\mathrm{D}}_{i}\bar{\phi}^{J}\delta\phi^{I}\right)\nonumber \\
&  & +\frac{1}{4}\mathcal{K}^{2}a^{2}\frac{\tilde{\epsilon}'\tilde{\epsilon}}{\mathcal{H}}\left(\bar{\mathrm{D}}^{2}+\frac{3-\tilde{\epsilon}}{2}\mathcal{K}\right)^{-2}\zeta\,\bar{\mathrm{D}}^{i}\left(\bar{f}_{IJ}\bar{\mathrm{D}}_{i}\bar{\phi}^{J}\delta\phi^{I}\right),\label{L2_zf}
\end{eqnarray}
and
\begin{equation}
\mathcal{L}_{2}^{(\gamma\delta\phi)}=-a^{2}\gamma^{ij}\bar{f}_{IJ}\left(\bar{\mathrm{D}}_{i}\bar{\mathrm{D}}_{j}\bar{\phi}^{J}+\bar{\mathrm{D}}_{i}\bar{\phi}^{K}\bar{\mathrm{D}}_{j}\bar{\phi}^{K}\bar{\Gamma}_{KL}^{J}\right)\delta\phi^{I}.\label{L2_gf}
\end{equation}

\section{$\mathcal{C}$'s and $\mathcal{D}$'s factors}

Here we collect the expressions for $\mathcal{C}$'s and $\mathcal{D}$'s factors from various contributions.

\begin{eqnarray}
\mathcal{C}_{ij,kl}^{(\alpha)}\left(\bm{k}_{1},\bm{k}_{2},\bm{k}_{3}\right) & = & \left(\delta_{ik}\delta_{jl}+\delta_{jk}\delta_{il}\right)\alpha\left(\bm{k}_{3}\right),\label{calC_ijkl_a}
\end{eqnarray} 
\begin{eqnarray}
\mathcal{D}_{ij,kl}^{(\alpha)}\left(\bm{k}_{1},\bm{k}_{2},\bm{k}_{3}\right) & = & \frac{3}{16}\left(k_{3j}k_{3l}\delta_{ik}+k_{3j}k_{3k}\delta_{il}+k_{3i}k_{3l}\delta_{jk}+k_{3i}k_{3k}\delta_{jl}\right)\alpha\left(\bm{k}_{3}\right)\nonumber \\
&  & +\frac{3}{16}\left(\bm{k}_{1}\cdot\bm{k_{2}}\right)\left(\delta_{il}\delta_{jk}+\delta_{ik}\delta_{jl}\right)\alpha\left(\bm{k}_{3}\right).\label{calD_ijkl_a}
\end{eqnarray}

	\begin{eqnarray}
\mathcal{C}_{ij,kl}^{\left(\beta\right)}\left(\bm{k}_{1},\bm{k}_{2},\bm{k}_{3}\right) & = & -2\Big[k_{3j}k_{3l}\delta_{ik}+k_{3j}k_{3k}\delta_{il}+k_{3i}k_{3l}\delta_{jk}+k_{3i}k_{3k}\delta_{jl}\nonumber \\
&  & -\frac{2}{3}k_{3}^{2}\left(\delta_{il}\delta_{jk}+\delta_{ik}\delta_{jl}\right)\Big]\beta\left(\bm{k}_{3}\right),\label{calC_ijkl_b}
\end{eqnarray} 
\begin{eqnarray}
\mathcal{D}_{ij,kl}^{\left(\beta\right)}\left(\bm{k}_{1},\bm{k}_{2},\bm{k}_{3}\right) & = & \frac{1}{8}\Big[\left(k_{3}^{2}-\left(\bm{k}_{1}\cdot\bm{k}_{2}\right)\right)\left(k_{3j}k_{3l}\delta_{ik}+k_{3j}k_{3k}\delta_{il}+k_{3i}k_{3l}\delta_{jk}+k_{3i}k_{3k}\delta_{jl}\right)\nonumber \\
&  & +\left(\left(\bm{k}_{1}\cdot\bm{k}_{2}\right)k_{3}^{2}-\left(\bm{k}_{1}\cdot\bm{k}_{3}\right)\left(\bm{k}_{2}\cdot\bm{k}_{3}\right)\right)\left(\delta_{il}\delta_{jk}+\delta_{ik}\delta_{jl}\right)\Big]\beta\left(\bm{k}_{3}\right).\label{calD_ijkl_b}
\end{eqnarray}

\begin{eqnarray}
&  & \mathcal{C}_{ij,kl}^{\left(\xi\right)}\left(\bm{k}_{1},\bm{k}_{2},\bm{k}_{3}\right)\nonumber \\
& = & i\big[k_{3l}\left(\delta_{jk}\xi_{i}\left(\bm{k}_{3}\right)+\delta_{ik}\xi_{j}\left(\bm{k}_{3}\right)\right)+k_{3k}\left(\delta_{jl}\xi_{i}\left(\bm{k}_{3}\right)+\delta_{il}\xi_{j}\left(\bm{k}_{3}\right)\right)\nonumber \\
&  & \quad+k_{3j}\left(\delta_{il}\xi_{k}\left(\bm{k}_{3}\right)+\delta_{ik}\xi_{l}\left(\bm{k}_{3}\right)\right)+k_{3i}\left(\delta_{jl}\xi_{k}\left(\bm{k}_{3}\right)+\delta_{jk}\xi_{l}\left(\bm{k}_{3}\right)\right)\big], \label{calC_ijkl_xi}
\end{eqnarray}
\begin{eqnarray}
&  & \mathcal{D}_{ij,kl}^{\left(\xi\right)}\left(\bm{k}_{1},\bm{k}_{2},\bm{k}_{3}\right)\nonumber \\
& = & \frac{1}{32}i\big\{\left(2\left(\bm{k}_{1}\cdot\bm{k}_{2}\right)-k_{3}^{2}\right)\big[k_{3l}\left(\delta_{jk}\xi_{i}\left(\bm{k}_{3}\right)+\delta_{ik}\xi_{j}\left(\bm{k}_{3}\right)\right)+k_{3k}\left(\delta_{jl}\xi_{i}\left(\bm{k}_{3}\right)+\delta_{il}\xi_{j}\left(\bm{k}_{3}\right)\right)\nonumber \\
&  & \qquad+k_{3j}\left(\delta_{il}\xi_{k}\left(\bm{k}_{3}\right)+\delta_{ik}\xi_{l}\left(\bm{k}_{3}\right)\right)+k_{3i}\left(\delta_{jl}\xi_{k}\left(\bm{k}_{3}\right)+\delta_{jk}\xi_{l}\left(\bm{k}_{3}\right)\right)\big]\nonumber \\
&  & \quad +2\left(\delta_{il}\delta_{jk}+\delta_{ik}\delta_{jl}\right)\left[\left(\bm{k}_{2}\cdot\bm{k}_{3}\right)\left(\bm{k}_{1}\cdot\bm{\xi}\left(\bm{k}_{3}\right)\right)+\left(\bm{k}_{1}\cdot\bm{k}_{3}\right)\left(\bm{k}_{2}\cdot\bm{\xi}\left(\bm{k}_{3}\right)\right)\right]\big\}. \label{calD_ijkl_xi}
\end{eqnarray}

\begin{equation}
\mathcal{C}_{ij,kl}^{\left(\eta\right)}\left(\bm{k}_{1},\bm{k}_{2},\bm{k}_{3}\right)=\delta_{il}\eta_{jk}\left(\bm{k}_{3}\right)+\delta_{ik}\eta_{jl}\left(\bm{k}_{3}\right)+\delta_{jl}\eta_{ik}\left(\bm{k}_{3}\right)+\delta_{jk}\eta_{il}\left(\bm{k}_{3}\right), \label{calC_ijkl_eta}
\end{equation}
\begin{eqnarray}
&  & \mathcal{D}_{ij,kl}^{\left(\eta\right)}\left(\bm{k}_{1},\bm{k}_{2},\bm{k}_{3}\right)\nonumber \\
& = & \frac{1}{16}\left(\bm{k}_{1}\cdot\bm{k}_{2}+k_{3}^{2}\right)\left(\delta_{il}\eta_{jk}\left(\bm{k}_{3}\right)+\delta_{ik}\eta_{jl}\left(\bm{k}_{3}\right)+\delta_{jl}\eta_{ik}\left(\bm{k}_{3}\right)+\delta_{jk}\eta_{il}\left(\bm{k}_{3}\right)\right)\nonumber \\
&  & +\frac{1}{16}\left(\delta_{il}\delta_{jk}+\delta_{ik}\delta_{jl}\right)k_{1m}k_{2n}\eta_{mn}\left(\bm{k}_{3}\right). \label{calD_ijkl_eta}
\end{eqnarray}


\providecommand{\href}[2]{#2}\begingroup\raggedright\endgroup

\end{document}